\documentclass[11]{article}

\usepackage{amsmath}
\usepackage{amsthm}
\usepackage{thmtools}
\usepackage{amssymb}
\usepackage{bbm}
\usepackage{mathtools}
\usepackage{tikz}
\usepackage{MnSymbol}
\usepackage{lipsum,hyperref}
\usepackage{graphicx}
\usepackage{amsfonts}
\usepackage{xfrac}
\usepackage{authblk}
\usepackage{color,soul}
\usepackage{framed}
\usepackage{setspace}
\usepackage{dutchcal}
\usepackage{rotating}
\usepackage{pdflscape}
\usepackage{setspace}
\usepackage{titlesec}
\usepackage[center]{caption}
\usepackage{subcaption}
\usepackage{xstring}
\usepackage{apxproof}
\usepackage{footnote}
\usepackage{etoolbox}
\usepackage{multirow}
\usepackage[framemethod=TikZ]{mdframed}
\usepackage[a4paper, margin=1in]{geometry}
\usetikzlibrary{matrix,patterns,decorations.pathreplacing,calc,fit}

\graphicspath{{./drawings/}}

\hypersetup{colorlinks=true}	

\newcommand{\eqdef}{\stackrel{\text{def}}{=}}

\newcommand{\set}[2]{\left\{ #1 \mathrel{}:\mathrel{} #2\right\}}

\newtheorem*{corollary*}{Corollary}
\newtheorem*{definition*}{Definition}

\DeclareMathOperator{\mon}{mon}

\DeclareMathOperator{\sens}{sens}

\DeclareMathOperator{\poly}{poly}
\DeclareMathOperator{\polylog}{polylog}

\DeclareMathOperator{\demand}{Demand}

\newcommand{\norm}[1]{\left\lVert#1\right\rVert}

\newcommand{\fouriernorm}[1]{\hat{\lVert}#1\hat{\rVert}}

\newcommand{\adeg}[1]{\widetilde{\deg}\left( #1 \right)}
\newcommand{\thirdadeg}[1]{\widetilde{\deg}_{\sfrac{1}{3}}\left( #1 \right)}
\newcommand{\epsadeg}[1]{\widetilde{\deg}_{\epsilon}\left( #1 \right)}

\DeclareMathOperator{\BPM}{BPM}
\DeclareMathOperator{\BPMn}{BPM_n}
\DeclareMathOperator{\BPMnstar}{BPM_n^\star}

\DeclareMathOperator{\DISJ}{DISJ}
\DeclareMathOperator{\AND}{AND}
\DeclareMathOperator{\ANDS}{AND_S}
\DeclareMathOperator{\ANDG}{AND_G}
\DeclareMathOperator{\ANDn}{AND_n}
\DeclareMathOperator{\OR}{OR}
\DeclareMathOperator{\ORn}{OR_n}

\DeclareMathOperator{\bs}{bs}

\newcommand\restr[2]{{\left.\kern-\nulldelimiterspace#1\vphantom{\big|}\right|_{#2}}}

\definecolor{nicegrey}{HTML}{e6e6ea}
\renewenvironment{leftbar}{%
	\MakeFramed {\advance\hsize-\width \FrameRestore}}%
{\endMakeFramed}

{\endMakeFramed}

\definecolor{niceorange}{HTML}{fed766}
\newenvironment{theo}[2][]{%
	\refstepcounter{theo}
	\ifstrempty{#1}%
	{\mdfsetup{%
			frametitle={%
				\tikz[baseline=(current bounding box.east),outer sep=0pt]
				\node[anchor=east,rectangle,fill=blue!20]
				{\strut Theorem~\thetheo};}
		}%
	}{\mdfsetup{%
			frametitle={%
				\tikz[baseline=(current bounding box.east),outer sep=0pt, inner ysep=1pt]
				\node[anchor=east,rectangle, draw=nicegrey, line width=0.7pt, fill=white]
				{\strut Theorem~\thetheo:~#1};}%
		}%
	}%
	\mdfsetup{%
		innertopmargin=1pt,linecolor=nicegrey,%
		linewidth=0.7pt,topline=true,%
		frametitleaboveskip=\dimexpr-\ht\strutbox\relax%
	}
	\begin{mdframed}[]\relax}{%
\end{mdframed}}

\definecolor{niceblue}{HTML}{0e9aa7}
\newenvironment{openq}[2][]{%
	\refstepcounter{openq}
	\ifstrempty{#1}%
	{\mdfsetup{%
			frametitle={%
				\tikz[baseline=(current bounding box.east),outer sep=0pt]
				\node[anchor=east,rectangle,fill=niceblue!20]
				{\strut Open Question~\theopenq};}
		}%
	}{\mdfsetup{%
			frametitle={%
				\tikz[baseline=(current bounding box.east),outer sep=0pt, inner ysep=1pt]
				\node[anchor=east,rectangle, fill=niceblue!15]
				{\strut Open Question~\theopenq:~#1};}%
		}%
	}%
	\mdfsetup{%
		innertopmargin=1pt,linecolor=niceblue!15,%
		linewidth=1pt,topline=true,%
		frametitleaboveskip=\dimexpr-\ht\strutbox\relax%
	}
	\begin{mdframed}[]\relax}{%
\end{mdframed}}

\newcommand{\keepvalues}{%
	\edef\restorevalues{%
		\parindent=\the\parindent
		\parskip=\the\parskip
	}%
}

\newtheorem{theorem}{Theorem}[section]
\newtheorem{corollary}{Corollary}[theorem]
\newtheorem{lemma}[theorem]{Lemma}
\newtheorem{proposition}[theorem]{Proposition}
\newtheorem{notation}[theorem]{Notation}
\newtheorem{definition}[theorem]{Definition}
\newtheorem{fact}[theorem]{Fact}

\newcounter{theo}[section]
\renewcommand{\thetheo}{\arabic{theo}}

\newcounter{openq}[section]
\renewcommand{\theopenq}{\arabic{openq}}
\newtheorem{openprob}[openq]{Open Problem}

\title{The Approximate Degree of Bipartite Perfect Matching}

\author{Gal Beniamini \\ \small{gal.beniamini@mail.huji.ac.il}} 
\affil{The Hebrew University of Jerusalem}

\begin{document}

\newcommand{\pad}{\phantom{0}}

\pgfkeys{tikz/adjmatrixenv/.style={
		decoration=brace,
		every left delimiter/.style={xshift=4.7pt},
		every right delimiter/.style={xshift=-4.7pt}}}
	
\pgfkeys{tikz/adjmatrix/.style={matrix of math nodes,
								nodes in empty cells,
								left delimiter=(,right delimiter={)},
								inner sep=2pt,
								column sep=1em,
								row sep=0.5em,
								nodes={inner sep=0pt}}}
							
\pgfkeys{tikz/adjmatrixbrace/.style={decorate,thick}}

\makeatletter
\def\tikz@lib@fit@scan{%
	\pgfutil@ifnextchar\pgf@stop{\pgfutil@gobble}{%
		\pgfutil@ifnextchar\foreach{\tikz@lib@fit@scan@handle@foreach}{%
			\tikz@scan@one@point\tikz@lib@fit@scan@handle}}}
\def\tikz@lib@fit@scan@handle@foreach\foreach#1in#2#3{%
	\foreach #1 in {#2}
	{\tikz@scan@one@point\tikz@lib@fit@scan@handle@foreach@#3}
	\tikz@lib@fit@scan}
\def\tikz@lib@fit@scan@handle@foreach@#1{%
	\iftikz@shapeborder
	\tikz@lib@fit@adjust{%
		\pgfpointanchor{\tikz@shapeborder@name}{west}}%
	\tikz@lib@fit@adjust{%
		\pgfpointanchor{\tikz@shapeborder@name}{east}}%
	\tikz@lib@fit@adjust{%
		\pgfpointanchor{\tikz@shapeborder@name}{north}}%
	\tikz@lib@fit@adjust{%
		\pgfpointanchor{\tikz@shapeborder@name}{south}}%
	\else
	\tikz@lib@fit@adjust{#1}%
	\fi
	\global\pgf@xa=\pgf@xa
	\global\pgf@ya=\pgf@ya
	\global\pgf@xb=\pgf@xb
	\global\pgf@yb=\pgf@yb}
\makeatletter

\pgfdeclarepatternformonly{north east lines b}{\pgfqpoint{0pt}{0pt}}{\pgfqpoint{20pt}{20pt}}{\pgfqpoint{20pt}{20pt}}%
{
	\pgfsetlinewidth{6pt}
	\pgfpathmoveto{\pgfqpoint{0pt}{0pt}}
	\pgfpathlineto{\pgfqpoint{20pt}{20pt}}
	\pgfpathmoveto{\pgfqpoint{16pt}{-4pt}}
	\pgfpathlineto{\pgfqpoint{24pt}{4pt}}
	\pgfpathmoveto{\pgfqpoint{-4pt}{16pt}}
	\pgfpathlineto{\pgfqpoint{4pt}{24pt}}
	\pgfusepath{stroke}
}

\newcommand*\adjmatrixbracebottom[4][m]{
	\draw[adjmatrixbrace, decoration={
		brace,
		raise=0.8em}]
	(#1.south-|#1-1-#2.north east) -- node[below=2pt, yshift=-0.8em] {#4} (#1.south-|#1-1-#3.north west);
}

\newcommand*\adjmatrixbracetop[4][m]{
	\draw[adjmatrixbrace] (#1.north-|#1-1-#2.south east) -- node[above=3pt] {#4} (#1.north-|#1-1-#3.south west);
}

\newcommand*\adjmatrixbraceright[4][m]{
	\draw[adjmatrixbrace, decoration={
		brace,
		mirror,
		raise=3mm}]
	(#1.east|-#1-#3-1.south east) -- node[right=3pt, xshift=3mm] {#4} (#1.east|-#1-#2-1.north east);
}

\maketitle

\begin{abstract}
	The approximate degree of a Boolean function is the least degree of a real multilinear polynomial approximating it in the $\ell_\infty$-norm over the Boolean hypercube. We show that the approximate degree of the Bipartite Perfect Matching function, which is the indicator over all bipartite graphs having a perfect matching, is $\widetilde{\Theta}(n^{\sfrac{3}{2}})$.
	
	The upper bound is obtained by fully characterizing the unique multilinear polynomial representing the Boolean dual of the perfect matching function, over the reals. Crucially, we show that this polynomial has very small $\ell_1$-norm -- only exponential in $\Theta(n \log n)$. The lower bound follows by bounding the spectral sensitivity of the perfect matching function, which is the spectral radius of its cut-graph on the hypercube \cite{aaronson2020degree, huang2019induced}. We show that the spectral sensitivity of perfect matching is exactly $\Theta(n^{\sfrac{3}{2}})$.
\end{abstract}

\pagebreak

\tableofcontents

\pagebreak

\section{Introduction}

The approximate degree of a Boolean function is the least degree of a real polynomial approximating the function in the $\ell_\infty$-norm over the Boolean cube, to within constant error. Approximate degree is an important complexity measure with applications throughout theoretical computer science. Lower bounds on the approximate degree of a Boolean function imply such bounds on the communication complexity (of a related composed problem) \cite{sherstov2011pattern, shi2007quantum}, and for its quantum query complexity \cite{beals2001quantum}. For families of Boolean functions, upper bounds on the approximate degree have algorithmic merit, for instance in learning theory \cite{klivans2004learning, kalai2008agnostically} and differential privacy \cite{thaler2012faster, chandrasekaran2014faster}, and conversely lower bounds imply separations in circuit complexity \cite{minsky1988perceptrons, sherstov2009separating}. For a recent survey, we refer the reader to \cite{bun2021guest}.

In this paper we study the approximate degree of the bipartite perfect matching function. This is the Boolean function representing the \textit{decision problem} of perfect matching --  determining whether a given balanced bipartite graph contains a subset of edges in which every vertex is incident to exactly one edge.

\begin{leftbar}
\begin{definition*}
	The bipartite perfect matching function $\BPMn: \{0,1\}^{n^2} \rightarrow \{0,1\}$ is defined as follows:
	\[ \BPMn(x_{1,1}, \dots, x_{n,n}) = \begin{cases*} 
	1\quad &$\set{(i,j)}{x_{i,j} = 1}$ \text{has a Bipartite Perfect Matching} \\
	0\quad &\text{Otherwise} 
	\end{cases*} \] 
\end{definition*}
\end{leftbar}

The input bits of $\BPMn$ select a subset of edges from the complete bipartite graph, and the output bit is set to $1$ if and only if the chosen subgraph contains a bipartite perfect matching of order $n$.

It is well known that any Boolean function can be uniquely and \textit{exactly} represented by a multilinear polynomial over the reals (see \cite{o2014analysis}). In \cite{beniamini2020bipartite}, the unique polynomial representing $\BPMn$ was characterized, and in particular was shown to have full degree, $n^2$. Conversely, it is not hard to construct low-degree polynomials \textit{approximating} the perfect matching function, if one allows pointwise errors arbitrarily close to one half. Indeed, the $n \times n$ Permanent implies (by translation and scaling) such a polynomial of total degree $n$, and approximation error exponentially close to half. Approximate degree is an interpolation between these two settings, wherein we require the errors be bounded by an arbitrary constant \textit{less than half}, say one third. The previous best-known upper bound on the approximate degree of perfect matching was $\mathcal{O}(n^{\sfrac{7}{4}})$, due to Lin and Lin \cite{lin2015upper}, and no non-trivial lower bound was known.

Our main result is the following approximate degree bound\footnote{In fact, we show that the same bound also holds even for approximations with \textit{exponentially small} error, see Section \ref{section:upper_bound}.}, which is tight up to low order terms. 

\begin{theo}[The Approximate Degree Bipartite Perfect Matching] {thm:apx_deg_bpm}
	\label{thm:apx_deg_bpm}
	For every $n \in \mathbb{N}$, the approximate degree of the bipartite perfect matching function is:
	\[\adeg{\BPMn} = \widetilde{\Theta}\left(n^{\sfrac{3}{2}}\right)\]
\end{theo}

Most known techniques for bounding approximate degree are applicable only to functions which are either \textit{symmetric} or \textit{block-composed} (with some recent notable exceptions, e.g. \cite{bun2018polynomial, bun2019nearly}). The perfect matching function falls into neither category, and is thus not amenable to standard techniques. 

Our upper bound follows by investigating the ``Boolean Dual'' function of bipartite perfect matching: $\BPMnstar(x_{1,1}, \dots, x_{n,n}) = 1 - \BPMn(1-x_{1,1}, \dots, 1-x_{n,n})$. In this representation, we reverse the roles of the symbols $0$ and $1$. Concretely, for any input graph, the dual function $\BPMnstar$ outputs $1$ if and only if the \textit{complement} of the graph \textit{does not} contain a bipartite perfect matching. Equivalently, by Hall's Marriage Theorem, the output is $1$ if and only if the input graph \textit{contains} a biclique over $n+1$ vertices. 

To present our characterization of the dual, let us introduce some notation. A balanced bipartite graph is said to be \textit{totally ordered}, if there exists an ordering of its left vertices such that their neighbour sets form a chain with respect to inclusion, i.e. $N(a_1) \subseteq N(a_2) \subseteq \dots \subseteq N(a_n)$. We associate with every totally ordered graph a ``representing sequence'', which encodes its biadjacency matrix up to permutations over both bipartitions. To construct this sequence, consider the automorphism which sorts the left and right vertices in descending order of degree. This yields a graph whose biadjacency matrix consists of a monotonically increasing sequence of \textit{blocks}, which we succinctly describe using a list of pairs of integers, describing the width and height of each such block. By way of example, the biclique $K_{s,t} \subseteq K_{n,n}$ is an ordered graph whose biadjacency matrix consists of two blocks; the first $s$ left vertices are all adjacent to the first $t$ vertices on the right, and the remainder are all isolated.


\pagebreak 

Our result is the following \textit{complete characterization} of the unique polynomial representing $\BPMnstar$ over the reals, thereby resolving an open question of \cite{beniamini2020bipartite}.

\begin{theo}[The Dual Polynomial of Bipartite Perfect Matching]{thm:bpm_star_poly}
	\label{thm:bpm_star_poly}
	
	\[\BPMnstar(x_{1,1}, \dots, x_{n,n}) = \sum_{G \subseteq K_{n,n}} a^\star_G \prod_{(i,j) \in E(G)} x_{i,j}\]
	
	\begin{itemize}
		\item If $G$ is not totally ordered, then $a^\star_G = 0$.
		\item Otherwise:
		\begin{equation*}
		\begin{split}
		a^\star_G = {{n-k_{t-1}-1} \choose {n - d_t}} \cdot \prod_{i=1}^{t-1} f\left(d_{i+1} - k_{i-1}, d_i - k_{i-1}, k_i - k_{i-1}\right)
		\end{split}
		\end{equation*}
		where $0 \le d_1 < d_2 < \dots < d_t \le n$ and $0 = k_0 < k_1 < k_2 < \dots < k_t = n$ form the \textit{representing sequence} of $G$, and the function $f: \mathbb{Z}^3 \rightarrow \mathbb{Z}$ is defined:
		\begin{equation*}
		\begin{split}
		f(n,d,k) = \begin{dcases*}
		{{n-1} \choose {k}}, & $d \le 0$ \\
		- {{n-d-1} \choose {k-d}} {{k-1} \choose {d-1}}, & $d>0$
		\end{dcases*}
		\end{split}
		\end{equation*}
	\end{itemize}
	
\end{theo}

This characterization allows us to deduce that the $\ell_1$-norm of $\BPMnstar$ (i.e., the sum of the magnitudes of its coefficients) is very small -- only exponential in $\Theta(n \log n)$. The approximate degree upper bound then follows via two observations. Firstly, we relate the approximate degree of any Boolean function and its dual. Secondly, we show that any Boolean function whose representation over the $\{0,1\}$-basis has low $\ell_1$-norm, can be efficiently approximated in the $\ell_\infty$-norm. The latter approach had also previously been employed by Sherstov in \cite{sherstov2020algorithmic}. 

To obtain the lower bound on the approximate degree of matching, we consider a new complexity measure recently introduced by Aaronson, Ben-David, Kothari, Rao and Tal \cite{aaronson2020degree}. For any total Boolean function $f$, they define the \textit{Spectral Sensitivity} to be the spectral radius of the bipartite graph defined by the $f$-bichromatic edges of the Hypercube (i.e., the $f$-cut of the cube). The notion of spectral sensitivity had notably (implicitly) also appeared at the heart of Huang's breakthrough proof of the Sensitivity Conjecture \cite{huang2019induced}. The main technical Theorem of \cite{aaronson2020degree} states that the spectral sensitivity of any total Boolean function lower bounds its approximate polynomial degree -- and it is this relation that we leverage.

We prove the following tight bound on the spectral sensitivity of $\BPMn$.

\begin{theo}[The Spectral Sensitivity of Bipartite Perfect Matching]{thm:spec_sens_bpm_n}
	\label{thm:spec_sens_bpm_n}
	The Spectral Sensitivity of the bipartite perfect matching function is $\lambda(\BPMn) = \Theta(n^{\sfrac{3}{2}})$.  
\end{theo}

One of our main motivations in studying the algebraic properties of $\BPMn$ and its dual, is the following longstanding question: what is the least complexity of a \textit{deterministic} algorithm for bipartite matching? Hopcroft and Karp's algorithm \cite{hopcroft1973n} from half a century ago attains a running time of $\mathcal{O}\left(n^{\sfrac{5}{2}}\right)$\footnote{On \textit{dense} graphs, wherein the number of edges is proportional to $n^2$.}, and as of yet no known deterministic algorithm has been shown to break the ``$n^{\sfrac{5}{2}}$-barrier''. In the last section of this paper we explore the above barrier through the lens of the Demand Query Model \cite{nisan2021demand}, which is a concrete complexity model for matching due to Nisan. The demand model was shown in \cite{nisan2021demand} to ``capture'' the complexity of a wide class of algorithms (i.e., \textit{combinatorial} algorithms), therefore any non-trivial lower bound on algorithms within the model would have far reaching implications. To this end, we draw connections between the \textit{algebraic quantities} explored throughout this work, including approximate degree and the $\ell_1$-norm of the dual, and the \textit{demand query complexity} of matching -- see Figure \ref{fig:complexity_measures}. Furthermore, we exhibit an efficient quantum simulation for the demand model, showing that lower bounds in the quantum query model yield corresponding combinatorial bounds. The quantum query complexity of matching was shown by Zhang \cite{zhang2004power} to be at least $\Omega\left(n^{\sfrac{3}{2}}\right)$, and by Lin and Lin \cite{lin2015upper} to be at most $\mathcal{O}\left(n^{\sfrac{7}{4}}\right)$. Closing this gap is left as an open question, and we remark that any polynomial improvement on the lower bound would yield a non-trivial bound in the demand model\footnote{Theorem \ref{thm:apx_deg_bpm} implies that this lower bound \textit{cannot be (polynomially) strengthened} by the ``polynomial method'', and in fact it is known that neither can Ambainis' adversary bounds be used to this end, see \cite{zhang2004power}.}. Finally, we remark that all the bounds obtained in this paper are compatible with the existence of quasi-linear demand query algorithms for bipartite matching, and this might be seen as weak evidence pointing in this direction. Obtaining non-trivial bounds on the demand query complexity of matching is left as our main open problem.

\subsection{Related Work}

The main objects of study in this paper are the perfect matching function and its dual. These functions had previously played a central role in \cite{beniamini2020bipartite}, wherein a complete characterization of $\BPMn$ was given, alongside a \textit{partial} description of the \textit{support} (i.e., non-zero coefficients) of $\BPMnstar$. In this paper we obtain a complete closed-form characterization of the dual function (see Theorem \ref{thm:bpm_star_poly}). To contextualize our result, let us briefly describe the techniques prior to this paper.

The unique polynomial representing $\BPMn$ was shown in \cite{beniamini2020bipartite} to be intimately related to the \textit{face lattice} of the Birkhoff polytope (a well-known polytope, which is the convex hull of all $n \times n$ permutation matrices). In particular, it was shown that every monomial corresponds to a \textit{face} of this polytope, and every coefficient is given by the M\"obius number of that face. A key component in the proof of this fact was a theorem due to Billera and Sarangarajan \cite{billera1994combinatorics}, stating that the lattice of all ``Matching-Covered Graphs'' (see Subsection \ref{subsection:matching_covered_elementary}) is isomorphic to the face lattice of the Birkhoff polytope.

The partial characterization of the dual polynomial, obtained in the same paper, \textit{heavily relied on the aforementioned lattice-theoretic approach}. In particular, the proof leveraged the fact that the matching-covered lattice is ``Eulerian'' (see \cite{Stanley:2011:ECV:2124415}). This observation sufficed in order to show that, for the vast majority of graphs, a certain degeneracy condition holds, implying that their corresponding dual coefficient must vanish.

While the lattice-based approach was fruitful in restricting the support of the dual, it was rather \textit{coarse} and therefore not well suited to proving a fine-grained characterizations, such as that of Theorem \ref{thm:bpm_star_poly}. The main problem occurs for graphs wherein the degeneracy condition above \textit{does not hold}, in which case the dual coefficient does not necessarily vanish. For every such graph, the technique above allows one to express its dual coefficient as a sum over a set of primal coefficients, however the number of summands is typically \textit{exponentially large} (in $n^2$). Since every term in this sum is either $\pm 1$, one would expect many cancellations to occur -- understanding these cancellations fully is one of the main challenges.\footnote{For our upper bound on the approximate degree of $\BPMn$, we require that the magnitude of all coefficients in the dual polynomial be at most exponential in $\Theta(n \log n)$. When expressed as a sum over $2^{\Theta(n^2)}$ summands, it is not immediately obvious why that should be the case. Indeed, by analogy to the ``drunkard's walk'' on the integers, one might expect such sums to be at most proportional to the square root of the number of steps taken, which is still exponential in $n^2$. Nevertheless, we show that these sums are far more regular and ``well-behaved'', and in fact are \textit{only} ever at most exponential in $2n$.}

To obtain a closed-form expression for the dual coefficients, we take a more combinatorial approach. Firstly, we show that with regards to the dual polynomial, one need only consider the \textit{connected components} of matching-covered graphs, known as ``Elementary Graphs''. The combinatorial properties of this family are well understood \cite{lovasz1979determinants, hetyei1964rectangular}, and play a key role in our proof. Secondly, we observe that the set of graphs appearing in the dual polynomial admit a decomposition into simpler components, which we dub ``blocks''. By doing so, we reduce the computation of any dual coefficient of an arbitrary graph to a product of dual coefficients corresponding to blocks. The decomposition scheme is rather involved, and is detailed fully in Section \ref{section:dual_polynomial}.
%

Finally, let us remark that our results regarding the dual polynomial can be similarly cast in a lattice-theoretic fashion. Much in the same way that $\BPMn$ was shown to be related to the matching-covered lattice, analogously $\BPMnstar$ can be shown to be related to the lattice of graphs covered by ``Hall-Violators'' (i.e., bicliques over $n+1$ vertices). In this light, Theorem \ref{thm:bpm_star_poly} can be viewed as a description of the M\"obius function of the aforementioned lattice.

\pagebreak
\section{Preliminaries and Notation}

\subsection{Boolean Functions and Polynomial Representation}

Let $f:\ \{0,1\}^n \rightarrow \{0,1\}$ be a Boolean function. The polynomial $p \in \mathbb{R}[x_1, \dots, x_n]$ \textit{represents} $f$, if for every $x \in \{0,1\}^n$, we have $p(x) = f(x)$. Recall the following useful fact regarding Boolean functions: 

\begin{leftbar}
	\begin{fact}
		Any Boolean function $f:\ \{0,1\}^n \rightarrow \{0,1\}$ can be \underline{uniquely} represented by a multilinear polynomial over the reals.
	\end{fact}
\end{leftbar}

Given the unique multilinear polynomial $p(x_1, \dots, x_n) = \sum_{S \subseteq [n]} a_S \left(\prod_{i \in S} x_i\right)$, representing a Boolean function $f: \{0,1\}^n \rightarrow \{0,1\}$, we denote by $\mon(f) \eqdef \set{S \subseteq [n]}{a_S \ne 0}$ the set of all monomials appearing in the polynomial representing $f$. Furthermore, we define the following two ``norms'', which are defined using the unique representations of Boolean functions, over the Boolean and Fourier bases.

\begin{leftbar}
	\begin{definition}
		Let $f: \{0,1\}^n \rightarrow \{0,1\}$ be a Boolean function. Let $p(x_1, \dots, x_n) = \sum_{S \subseteq [n]} a_S \Pi_{i \in S} x_i$ be the unique multilinear polynomial representing $f$ over the reals, and let $\{\widehat{f}_S :S \subseteq [n]\}$ be the Fourier spectrum of $f$. The $\ell_1$-norm of $f$ is defined:
		
		\[ \norm{f}_1 \eqdef \norm{p}_1 \eqdef \sum_{S \subseteq [n]} |a_S| \]
		and similarly, the Fourier $\ell_1$-norm of $f$ is defined:
		\[ \fouriernorm{f}_1 \eqdef \sum_{S \subseteq [n]} |\widehat{f}_S| \]
	\end{definition}
\end{leftbar}

The $\epsilon$-approximate degree of a Boolean function $f: \{0,1\}^n \rightarrow \{0,1\}$ is the least degree of a real multilinear polynomial \textit{approximating} $f$ in the $\ell_\infty$ norm, with error at most $\epsilon$. Hereafter, we use the standard notation and write $\epsadeg{f}$ to denote the $\epsilon$-approximate degree of $f$. In the case of $\epsilon=\tfrac{1}{3}$, we omit the $\epsilon$ and instead write $\adeg{f}$.

\begin{leftbar}
	\begin{definition}
		Let $f: \{0,1\}^n \rightarrow \{0,1\}$ be a Boolean function, and let $0 < \epsilon < \tfrac{1}{2}$. The $\epsilon$-approximate degree of $f$, $\epsadeg{f}$, is the least degree of a Real polynomial $p \in \mathbb{R}[x_1, \dots, x_n]$ such that:
		\[ \forall x \in \{0,1\}^n:\ |f(x) - p(x)| \le \epsilon \] 
	\end{definition}
\end{leftbar} 

In the context of Boolean functions, it is sometimes useful to consider the transformation of a Boolean vector in which an arbitrary subset of bits have been flipped. Thus, if $x \in \{0,1\}^n$, and $S \subseteq [n]$, we use the notation $x^S$ to indicate the vector in which the coordinates $S$ have been flipped. For any $i \in [n]$, the notation $x^i$ is shorthand for $x^{\{i\}}$. Using this notation, we define the following two complexity measures for Boolean functions.

\begin{leftbar}
	\begin{definition}
		\label{def:sensitivity_and_block_sensitivity}
		Let $f\: \{0,1\}^n \rightarrow \{0,1\}$ be a Boolean function. The \textbf{sensitivity} of $f$ at $x \in \{0,1\}^n$ is:
		\[ \sens_f(x) = \left|\set{i \in [n]}{f(x) \ne f(x^{i})}\right| \]
		and similarly, the \textbf{block sensitivity}  at $x$ is:
		\[ \bs_f(x) = \max\{s \in [n], \text{ such that } \exists B_1 \cupdot \dots \cupdot B_s \subseteq [n]:\ \forall i \in [s]:\ f(x) \ne f(x^{B_i})\} \]
		
		The \textit{sensitivity} and \textit{block sensitivity} of $f$ are then defined by their corresponding measures on the worst case input, namely $\sens(f) = \max_{x \in \{0,1\}^n} \sens_f(x)$ and $\bs(f) = \max_{x \in \{0,1\}^n} \bs_f(x)$.
	\end{definition}
\end{leftbar}

\subsection{Graph Theory}

We use standard definitions and notation relating to graphs. If $G$ is a graph, we denote its vertex set by $V(G)$, its edge set by $E(G)$, and its connected components by $C(G)$. For any vertex $v \in V(G)$, the neighbour set of $v$ is denoted by $N(v)$, and its degree is denoted $\deg(v) = |N(v)|$. The set of all \textit{perfect matchings} of $G$ is denoted by $PM(G)$. We also use the following slightly less common quantity:

\begin{leftbar}
	\begin{definition}
		\label{def:cyclomatic_number}
		Let $G$ be a graph. The \textbf{cyclomatic number} of $G$ is defined by: 
		\[ \chi(G) = |E(G)| - |V(G)| + |C(G)| \]
	\end{definition}
\end{leftbar}

The graph $G - v$, where $v \in V(G)$, is the graph over the vertices $V(G) \setminus \{v\}$ in which all the edges incident to $v$ are omitted. If $U \subseteq V(G)$ is a set of vertices, the notation $G\left[U\right]$ refers to the \textit{induced graph} on the vertices $U$, whose vertices are $U$ and whose edges are the edges of $G$ which are incident only to vertices in $U$. If $G \subseteq K_{n,n}$ and $S \subseteq E(K_{n,n})$ the notation $G \cup S$ refers the a graph over the vertices of $K_{n,n}$, whose edge set is $E(G) \cup S$.

For any graph $G$, the \textit{adjacency matrix} $A_G$ is a symmetric matrix whose rows and columns are labeled by $V(G)$, and whose entries are given by $(A_G)_{u,v} = \mathbbm{1}\{ \{u,v\} \in E(G) \}$. The \textit{spectral radius} of $G$ is defined $\rho(G) \eqdef \max \lbrace |\lambda_i| : \lambda_i \in Spec(A_G)\rbrace$, i.e., the maximum magnitude of any eigenvalue in the spectrum of $A_G$. Since the spectrum of bipartite graphs is \textit{symmetric}, it holds that for any bipartite graph $\rho(G) = \lambda_1$.

Throughout this paper, we restrict our attention to \textit{balanced bipartite graphs over the vertices of the complete bipartite graph}, $K_{n,n}$. By convention, we label the left vertices of $K_{n,n}$ by $a_1, \dots, a_n$, and the right vertices by $b_1, \dots, b_n$. The notation $G \subseteq K_{n,n}$ is used to indicate that $G$ is a balanced bipartite graph over the vertices of $K_{n,n}$. Similarly, the notation $G \subseteq H$ indicates that $V(G)=V(H)$ and $E(G) \subseteq E(H)$.

\subsection{Quantum Query Complexity}

We consider the standard quantum query model (see, e.g., \cite{buhrman2002complexity}). For a recent textbook on the framework of quantum computing, we refer the reader to \cite{nielsen2002quantum}. In this paper, we refer to the bounded-error quantum query complexity, which is defined as follows:

%
%

\begin{leftbar}
	\begin{definition}
		Let $f: \{0,1\}^n \rightarrow \{0,1\}$ be a Boolean function. The bounded-error quantum query complexity of $f$, $Q_2(f)$, is the smallest number $d$, such that there exists a quantum query algorithm $\mathcal{A}$ making at most $d$ queries, and satisfying:
		
		\[ \forall x \in \{0,1\}^n: \Pr\left[ \mathcal{A}(x) = f(x) \right] \ge \tfrac{2}{3}  \]
	\end{definition}
\end{leftbar}
\pagebreak
\section{The Dual Polynomial of Bipartite Perfect Matching}
\label{section:dual_polynomial}

This section centers around the proof of Theorem \ref{thm:bpm_star_poly}. To provide the proof, we must first familiarize ourselves with some useful definitions and notation. To this end, we begin by defining Boolean dual functions, and by recalling two relevant graph families: matching-covered graphs, and elementary graphs. Then, we introduce the notion of sorted and ordered graphs, which serve as the building blocks of our proof. Finally, we provide our proof of Theorem \ref{thm:bpm_star_poly}. 

\subsection{Boolean Dual Functions}

\begin{leftbar}
	\begin{definition}
		Let $f: \{0,1\}^n \rightarrow \{0,1\}$ be a Boolean function. The \textbf{Boolean Dual} function of $f$ is denoted $f^\star: \{0,1\}^n \rightarrow \{0,1\}$ and is defined as follows:
		
		\[ f^\star(x_1, \dots, x_n) = 1 - f(1-x_1, \dots, 1-x_n) \]
		
	\end{definition}
\end{leftbar}

Intuitively, in the Boolean dual, the symbols $0$ and $1$ switch roles. Geometrically, if we consider $f$ to be a colouring of the vertices of the $n$-dimensional hypercube, the duality transformation simply mirrors all vertices and inverts their colours. Algebraically, when representing the functions using multilinear polynomials over the reals, each monomial in the ``primal'' function corresponds to an $\AND$ function, whereas in the dual each monomial corresponds to an $\OR$ (over the original input bits). A Boolean function $f$ and its dual $f^\star$ share many properties. For example, their Fourier spectra are identical (up to signs of Fourier coefficients, see \cite{o2014analysis}). Nevertheless, in the $\{0,1\}$ basis, the unique multilinear polynomials representing $f$ and $f^\star$ can differ greatly. By way of example, the polynomial representing $\ANDn$ consists of a single monomial, whereas its dual ($\AND^\star_n = \ORn$) has exactly $2^n - 1$ monomials. 

\subsection{Matching-Covered and Elementary Graphs}
\label{subsection:matching_covered_elementary}

A graph $G \subseteq K_{n,n}$ is said to be \textit{matching-covered} if every edge of $G$ participates in some perfect matching, or equivalently if its edge set can be described as the union over a set of perfect matchings $S \subseteq PM(G)$.

\begin{leftbar}
	\begin{definition}
		\label{def:matching_covered_graph}
		Let $G \subseteq K_{n,n}$ be a graph. $G$ is \textbf{matching-covered} if and only if:
			\[ \forall e \in E(G):\ \exists M \in PM(G):\ e \in M \] 
	\end{definition}
\end{leftbar}

%
Matching-covered graphs have many interesting combinatorial properties. The set of all such graphs, together with the subset relation over the edges, forms a lattice. A key result by Billera and Sarangarajan \cite{billera1994combinatorics} showed that this lattice is, in fact, isomorphic to the face lattice of the Birkhoff polytope, $B_n$. This lattice was later shown by \cite{beniamini2020bipartite} to be intimately related to the multilinear polynomial representing the bipartite perfect matching function, $\BPMn$. Namely, the monomials of the polynomial are the elements of the lattice, and their coefficients are the M\"obius numbers of this lattice. \\

A closely related family of graphs are the \textit{Elementary Graphs}.

\begin{leftbar}
	\begin{definition}
		\label{def:elementary_graph}
		Let $G \subseteq K_{n,n}$ be a graph. Then:
		\[G \text{ is \textbf{elementary}} \iff G \text { is a \underline{connected} matching-covered graph} \]
	\end{definition}
\end{leftbar}

Hereafter, we denote all matching-covered graphs by $MC_n = \set{G \subseteq K_{n,n}}{G \text{ is matching-covered}}$, and similarly we denote all elementary graphs by $EL_n = \set{G \subseteq K_{n,n}}{G \text{ is elementary}}$. Elementary graphs were studied at length, both by Lov{\'a}sz and Plummer \cite{plummer1986matching}, and earlier by Hetyei \cite{hetyei1964rectangular}. Through their works they formulated robust characterizations of elementary graphs. In particular, we require the following useful theorem, due mostly to Hetyei:

\begin{leftbar}
	\begin{theorem}[\cite{hetyei1964rectangular}]
		\label{thm:hetyei_conditions}
		Let $G=(A \cupdot B, E)$ be a bipartite graph. The following are equivalent:
		\begin{itemize}
			\item $G$ is elementary.
			\item $G$ has exactly two minimum vertex covers, $A$ and $B$.
			\item $|A|=|B|$ and for every $\emptyset \ne X \subset A$, $|N(X)| > |X|$.
			\item $G=K_2$, or $|V(G)| \ge 4$ and for any $a \in A$, $b \in B$, $G - a - b$ has a perfect matching.
			\item $G$ is connected and every edge is ``\textit{allowed}'', i.e., appears in a perfect matching of $G$.
		\end{itemize}
	\end{theorem}
\end{leftbar}

\subsection{Ordered Graphs}

\begin{leftbar}
\begin{definition}
	Let $G\subseteq K_{n,n}$. $G$ is a \textbf{totally ordered graph}, if there exists an ordering $\pi \in S_n$ of its left vertices, such that:
		\[ N(a_{\pi(1)}) \subseteq N(a_{\pi(2)}) \subseteq \dots \subseteq N(a_{\pi(n)}) \]
\end{definition}
\end{leftbar}

Given a totally ordered graph $G$, we may permute the vertices in its left and right bipartitions (separately) so that both bipartitions are sorted in decreasing order of degree. This automorphism produces a graph $H \cong G$, which we refer to as a ``sorted ordered graph''. Our motivation in applying such a transformation is due to the fact that $\BPMnstar$ is invariant to permutations over its bipartitions. Thus, the dual coefficient of any ordered graph and its corresponding sorted ordered graph are identical. 

\begin{leftbar}
	\begin{definition}
		Let $G\subseteq K_{n,n}$. $G$ is a \textbf{sorted ordered graph} if:
		\[\deg(a_1) \le \deg(a_2) \le \dots \le \deg(a_n)\]

		and furthermore:
		\begin{equation*}
			\begin{split}
				\forall i \in [n]:\ N(a_i) = \{b_1, \dots, b_{\deg(a_i)}\} 
			\end{split}	
		\end{equation*}
	\end{definition}
\end{leftbar}

The adjacency relation of a sorted ordered graph can be succinctly and uniquely described by a short sequence of integers, which we dub the ``representing sequence'' of the graph.

\begin{leftbar}
\begin{definition}
	Let $G \subseteq K_{n,n}$ be a sorted ordered graph. The \textbf{representing sequence} of $G$ is defined by: $\mathcal{S}_G = \{(d_1, k_1), \dots, (d_t, k_t)\}$, where:
	
	\begin{equation*}
		\begin{split}
			0 \le d_1 < d_2 < \dots < d_t \le n \\
			0 < k_1 < k_2 < \dots < k_t = n
		\end{split}
	\end{equation*}
	
	and furthermore: 
	\begin{equation*}
		\begin{split}
			\forall i \in [n]:\ N(a_i) = \begin{cases*}
			\{b_1, \dots, b_{d_1}\}, & $0 < i \le k_1$    \\
			\{b_1, \dots, b_{d_2}\}, & $k_1 < i \le k_2$  \\
			\ \vdots & $\ \vdots$ \\
			\{b_1, \dots, b_{d_t}\}, & $k_{t-1} < i \le k_t$
			\end{cases*}
		\end{split}
	\end{equation*}
	
\end{definition}
\end{leftbar}

The \textit{representing sequence} $\mathcal{S}_G$ of a sorted ordered graph $G \subseteq K_{n,n}$ is essentially a ``compressed'' form of its degree sequence; each pair $(d_i, k_i)$ in the sequence indicates a run of $(k_i - k_{i-1})$ left vertices, all of whose neighbour sets are exactly $\{b_1, \dots, b_{d_i}\}$. Thus, the biadjacency matrix of $G$ is simply described in terms of $\mathcal{S}_G$, as shown in Figure \ref{fig:repr_seqeunce}. 

\pagebreak

\newcommand*\adjmatrixbraceleftI[4][m]{
    \draw[adjmatrixbrace, decoration={
   							brace,
   							raise=5mm}]
   	(#1.west|-#1-#3-1.south west) -- node[left=1pt, xshift=-5mm] {#4} (#1.west|-#1-#2-1.north west);
}

\newcommand*\adjmatrixbraceleftII[4][m]{
	\draw[adjmatrixbrace, decoration={
		brace,
		raise=-14mm}]
	(#1.west|-#1-#3-1.south west) -- node[left=0pt, xshift=14mm] {#4} (#1.west|-#1-#2-1.north west);
}

\newcommand*\adjmatrixbraceleftIII[4][m]{
	\draw[adjmatrixbrace, decoration={
		brace,
		raise=-34mm}]
	(#1.west|-#1-#3-1.south west) -- node[left=0pt, xshift=34mm] {#4} (#1.west|-#1-#2-1.north west);
}

\begin{figure}[h!]
\begin{center}
\begin{tikzpicture}[baseline=0cm,adjmatrixenv]
	\matrix [adjmatrix,outer ysep=0.7pt,inner sep=8pt, row sep=1em, column sep=1.4em] (m)  
	{
		\phantom{0} &   &   &   &   &   &   &    \\
		\phantom{0} &   &   &   &   &   &   &    \\
		\phantom{0} &   &   &   &   & 1 & 1 & 1  \\
		\phantom{0} &   &   &   &   & 1 & 1 & 1  \\
		\phantom{0} &   & 1 & 1 & 1 & 1 & 1 & 1  \\
		\phantom{0} &   & 1 & 1 & 1 & 1 & 1 & 1  \\
		1           & 1 & 1 & 1 & 1 & 1 & 1 & 1  \\
		1           & 1 & 1 & 1 & 1 & 1 & 1 & 1  \\
	};
	
	\node [fit= \foreach \X in {3,...,8}{(m-\X-6)}
				\foreach \X in {3,...,8}{(m-\X-7)}
				\foreach \X in {3,...,8}{(m-\X-8)}]
	[draw=niceorange, thick,inner sep=2.6pt] (fit-rect1) {};
	
	\node [fit= \foreach \X in {5,...,8}{(m-\X-3)}
				\foreach \X in {5,...,8}{(m-\X-4)}
				\foreach \X in {5,...,8}{(m-\X-5)}]
	[draw=niceorange, thick,inner sep=2.6pt] (fit-rect2) {};
	
	\node [fit= \foreach \X in {7,...,8}{(m-\X-1)}
				\foreach \X in {7,...,8}{(m-\X-2)}]
	[draw=niceorange, thick,inner sep=2.6pt] (fit-rect3) {};
	
	\node[align=center, fit=(m-8-1.south), yshift=-1em, xshift=-0.4em] {\tiny $a_1$};
	\node[align=center, fit=(m-8-2.south), yshift=-1em, xshift=-0.4em] {\tiny $a_2$};
	\node[align=center, fit=(m-8-3.south), yshift=-1em, xshift=-0.4em] {\tiny $a_3$};
	\node[align=center, fit=(m-8-4.south), yshift=-1em, xshift=-0.4em] {\tiny $a_4$};
	\node[align=center, fit=(m-8-5.south), yshift=-1em, xshift=-0.4em] {\tiny $a_5$};
	\node[align=center, fit=(m-8-6.south), yshift=-1em, xshift=-0.4em] {\tiny $a_6$};
	\node[align=center, fit=(m-8-7.south), yshift=-1em, xshift=-0.4em] {\tiny $a_7$};
	\node[align=center, fit=(m-8-8.south), yshift=-1em, xshift=-0.4em] {\tiny $a_8$};
	
	\node[align=center, fit=(m-8-1.west), xshift=-1.8em, yshift=-0.2em] {\tiny $b_1$};
	\node[align=center, fit=(m-7-1.west), xshift=-1.8em, yshift=-0.2em] {\tiny $b_2$};
	\node[align=center, fit=(m-6-1.west), xshift=-1.8em, yshift=-0.2em] {\tiny $b_3$};
	\node[align=center, fit=(m-5-1.west), xshift=-1.8em, yshift=-0.2em] {\tiny $b_4$};
	\node[align=center, fit=(m-4-1.west), xshift=-1.8em, yshift=-0.2em] {\tiny $b_5$};
	\node[align=center, fit=(m-3-1.west), xshift=-1.8em, yshift=-0.2em] {\tiny $b_6$};
	\node[align=center, fit=(m-2-1.west), xshift=-1.8em, yshift=-0.2em] {\tiny $b_7$};
	\node[align=center, fit=(m-1-1.west), xshift=-1.8em, yshift=-0.2em] {\tiny $b_8$};

	
	\adjmatrixbracebottom{2}{1}{$k_1$}
	\adjmatrixbracebottom{5}{3}{$k_2 - k_1$}
	\adjmatrixbracebottom{8}{6}{$k_3 - k_2$}
	
	\adjmatrixbraceleftI{7}{8}{$d_1$}
	\adjmatrixbraceleftII{5}{6}{$d_2 - d_1$}
	\adjmatrixbraceleftIII{3}{4}{$d_3 - d_2$}
	
	\adjmatrixbraceright{1}{8}{$n$}
	\adjmatrixbracetop{1}{8}{$n$}
\end{tikzpicture}
\caption{The biadjacency matrix of a sorted ordered graph $G$.\\ The \textit{representing sequence} of $G$ is $\mathcal{S}_G = \{ (d_1, k_1), (d_2, k_2), (d_3, k_3)\}$}
\label{fig:repr_seqeunce}
\end{center}
\end{figure} 

The building blocks in our proof of Theorem \ref{thm:bpm_star_poly} consist of particular family of simple sorted ordered graphs -- those whose representing sequence is of length exactly $2$. In other words, these are the graphs whose left vertices can be partitioned into two sets, those having full degree $n$, and those whose neighbour set is (the same) strict subset of the right vertices. This family also trivially includes all bicliques $K_{s,n}$. For this family of graphs, we introduce the following notation. 

\begin{leftbar}
	\begin{notation}
		Let $n, d, k \in \mathbb{N}$, where $0 \le d \le n$ and $0 < k < n$. The notation $\langle n,d,k \rangle$-block refers to the sorted ordered graph $G \subseteq K_{n,n}$, whose representing sequence is:
		
		\[\mathcal{S}_{\langle n,d,k \rangle} = \{(d,k), (n,n)\}\]
	\end{notation}
\end{leftbar}

\begin{figure}[h!]
	\begin{center}
	\begin{tikzpicture}[baseline=0cm,adjmatrixenv]
	\matrix [adjmatrix,outer ysep=0.7pt,inner sep=8pt, row sep=1em, column sep=1.4em] (m)  
	{
		\phantom{0} &   &   & 1 & 1  \\
		\phantom{0} &   &   & 1 & 1  \\
		\phantom{0} &   &   & 1 & 1  \\
		1           & 1 & 1 & 1 & 1  \\
		1           & 1 & 1 & 1 & 1  \\
	};
	
	\adjmatrixbracebottom{3}{1}{$k$}
	\adjmatrixbraceright{1}{5}{$n$}
	\adjmatrixbraceleftI{4}{5}{$d$}

	\node [fit= \foreach \X in {4,...,5}{(m-\X-1)}
				\foreach \X in {4,...,5}{(m-\X-2)}
				\foreach \X in {4,...,5}{(m-\X-3)}]
	[draw=niceorange, thick,inner sep=2.6pt] (rect1) {};
	
	\node [fit= \foreach \X in {1,...,5}{(m-\X-4)}
				\foreach \X in {1,...,5}{(m-\X-5)}]
	[draw=niceorange, thick,inner sep=2.6pt] (rect2) {};
	
	\end{tikzpicture}
	\caption{The biadjacency matrix of an $\langle n,d,k \rangle$-block}
	\end{center}
\end{figure}

\subsection{Proof of Theorem \ref{thm:bpm_star_poly}}

\begin{leftbar}
	\begin{definition}
		\label{def:bpm_star}
		Let $\BPMnstar: \{0,1\}^{n^2} \rightarrow \{0,1\}$ be the Boolean dual function of $\BPMn$, defined by:
		\[\BPMnstar(x_{1,1}, \dots, x_{n,n}) = \begin{dcases*}
		1, & $\set{(i,j)}{x_{i,j} = 0}$ \text{does \underline{not} have a bipartite perfect matching} \\
		0, & \text{Otherwise}
		\end{dcases*}\]		
	\end{definition}
\end{leftbar}

In \cite{beniamini2020bipartite}, a complete characterization of the multilinear polynomial representing $\BPMn$ over the reals was obtained, using a connection between the M\"obius function of the Birkhoff polytope's face lattice, and the cyclomatic numbers of matching-covered graphs. The polynomial representing $\BPMnstar$ may be similarly expressed through the M\"obius function of some lattice (that of graphs covered by ``Hall Violators'', i.e., bicliques over $n+1$ vertices). These representations allowed for a \textit{partial} description of the \textit{support} of $\BPMnstar$, which we require for our proof of Theorem \ref{thm:bpm_star_poly} and will therefore now recall. The first two lemmas restrict the \textit{support} of monomials in the dual polynomial to the set of \textit{totally ordered graphs}, which are \textit{not matching-covered}. 

%
%

\begin{leftbar}
	\begin{lemma}[\cite{beniamini2020bipartite}]
		\label{lem:non_tot_ord_zero_coeff}
		Let $G \subseteq K_{n,n}$. If $G$ is not totally ordered, then $a^\star_G = 0$.
	\end{lemma}
\end{leftbar}

\begin{leftbar}
	\begin{lemma}[\cite{beniamini2020bipartite}]
		\label{lem:mc_dual_coeff_zero}
		Let $G \subseteq K_{n,n}$. If $G \in MC_n$, then $a^\star_G = 0$.
	\end{lemma}
\end{leftbar}

The third lemma relates the M\"obius numbers of the lattice of matching-covered graphs, with the dual coefficients of any graph $G \subseteq K_{n,n}$, thereby giving a closed-form expression for computing the dual coefficients (albeit by summing over possibly exponentially many summands).

\begin{leftbar}
	\begin{lemma}[\cite{beniamini2020bipartite}]
		\label{lem:dual_coeff_upper_sum}
		Let $G \subseteq K_{n,n}$. The dual coefficient of $G$ is:
		
		\[ a^\star_G = (-1)^{|E(G)|+1} \sum_{\substack{H \supseteq G \\ H \in MC_n}} (-1)^{\chi(H)} \]
	\end{lemma}
\end{leftbar}
\begin{leftbar}
	\begin{corollary}
		\label{cor:dual_coeff_diff_sum}
		Let $G \subseteq K_{n,n}$ be a graph. If all the left vertices or all the right vertices of $G$ are in the same connected component. Then:

		\[a^\star_G = \sum_{\substack{G \subseteq H \subseteq K_{n,n} \\ H \in EL_n}} (-1)^{|E(H) \setminus E(G)|}\]
	\end{corollary}
\end{leftbar}
\begin{proof}
	Recall that every connected component of a matching-covered graph is elementary. Furthermore, elementary graphs are balanced. Thus, $G \subseteq H \in MC_n$ $\implies$ $H$ is elementary, and we have:
	\begin{equation*}
		\begin{split}
			a^\star_G &= (-1)^{|E(G)|+1} \sum_{\substack{H \supseteq G \\ H \in MC_n}} (-1)^{\chi(H)} \\
			&= (-1)^{|E(G)|+1} \sum_{\substack{H \supseteq G \\ H \in EL_n}} (-1)^{|E(H)|  -2n + 1} 
			= \sum_{\substack{H \supseteq G \\ H \in EL_n}} (-1)^{|E(H) \setminus E(G)|} \qedhere
		\end{split}
	\end{equation*}
\end{proof}

\subsubsection{Reducing to Permitted Edges}

We now make the following observation: if $G$ is a totally ordered graph whose coefficient we wish to compute using Lemma \ref{lem:dual_coeff_upper_sum}, then we may \textit{restrict} our attention to a particular subset of edges. Whereas Lemma \ref{lem:dual_coeff_upper_sum} mandates that we consider every possible ``completion'' of $G$ to a matching-covered graph, the following lemma shows that we can instead only consider completions which are confined to the set of ``permitted edges'' for $G$. 

\begin{leftbar}
	\begin{definition}
		\label{def:permitted_edges}
		Let $G \subseteq K_{n,n}$ be a sorted ordered graph, and let $\mathcal{S}_G = \{(d_1, k_1), \dots, (d_t, k_t)\}$ be its representing sequence. The set of \textbf{permitted edges} for $G$ is denoted $\mathcal{P}_G$, and is defined by:
		
		\begin{equation*}
			(a_i, b_j) \in \mathcal{P}_G \iff \begin{dcases*} 
				d_1 < j \le d_2 , & $0 < i \le k_1$   \\
				d_2 < j \le d_3 , & $k_1 < i \le k_2$ \\
				\ \vdots & $\ \vdots$ \\
				d_{t} < j \le n, & $k_{t-1} < i \le k_t$
			\end{dcases*} 
		\end{equation*}
	\end{definition}
\end{leftbar}

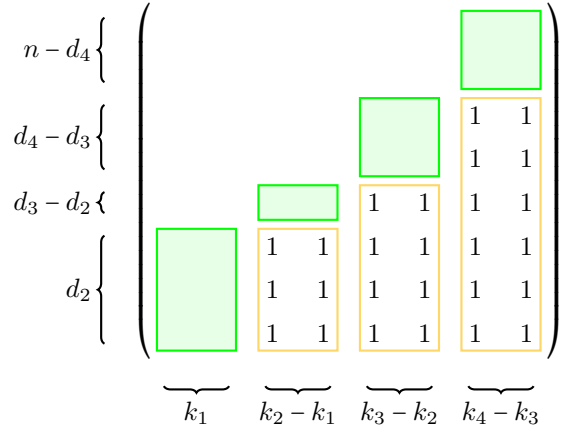
\begin{figure}[h!]
\begin{center}
	\begin{tikzpicture}[baseline=0cm,adjmatrixenv]
		\matrix [adjmatrix,outer ysep=0.7pt,inner sep=8pt, row sep=1em, column sep=1.4em] (m)  
		{
			\pad & \pad & \pad & \pad & \pad & \pad & \pad & \pad  \\
			\pad & \pad & \pad & \pad & \pad & \pad & \pad & \pad  \\
			\pad & \pad & \pad & \pad & \pad & \pad & 1    & 1     \\
			\pad & \pad & \pad & \pad & \pad & \pad & 1    & 1     \\
			\pad & \pad & \pad & \pad & 1    & 1    & 1    & 1     \\
			\pad & \pad & 1    & 1    & 1    & 1    & 1    & 1     \\
			\pad & \pad & 1    & 1    & 1    & 1    & 1    & 1     \\
			\pad & \pad & 1    & 1    & 1    & 1    & 1    & 1     \\
		};
		
		\node [fit= \foreach \X in {6,...,8}{(m-\X-1)}
					\foreach \X in {6,...,8}{(m-\X-2)}]
		[draw=green, fill=green, fill opacity=0.1, thick,inner sep=2.6pt] (per-rect1) {};
	
		\node [fit= \foreach \X in {5,...,5}{(m-\X-3)}
					\foreach \X in {5,...,5}{(m-\X-4)}]
		[draw=green, fill=green, fill opacity=0.1, thick,inner sep=2.6pt] (per-rect2) {};
	
		\node [fit= \foreach \X in {3,...,4}{(m-\X-5)}
					\foreach \X in {3,...,4}{(m-\X-6)}]
		[draw=green, fill=green, fill opacity=0.1, thick,inner sep=2.6pt] (per-rect3) {};
		
		\node [fit= \foreach \X in {1,...,2}{(m-\X-7)}
					\foreach \X in {1,...,2}{(m-\X-8)}]
		[draw=green, fill=green, fill opacity=0.1, thick,inner sep=2.6pt] (per-rect4) {};
			
		\node [fit= \foreach \X in {6,...,8}{(m-\X-3)}
					\foreach \X in {6,...,8}{(m-\X-4)}]
		[draw=niceorange, thick,inner sep=2.6pt] (block-rect1) {};
	
		\node [fit= \foreach \X in {5,...,8}{(m-\X-5)}
					\foreach \X in {5,...,8}{(m-\X-6)}]
		[draw=niceorange, thick,inner sep=2.6pt] (block-rect2) {};
	
		\node [fit= \foreach \X in {3,...,8}{(m-\X-7)}
					\foreach \X in {3,...,8}{(m-\X-8)}]
		[draw=niceorange, thick,inner sep=2.6pt] (block-rect3) {};

		\adjmatrixbracebottom{2}{1}{$k_1$}
		\adjmatrixbracebottom{4}{3}{$k_2 - k_1$}
		\adjmatrixbracebottom{6}{5}{$k_3 - k_2$}
		\adjmatrixbracebottom{8}{7}{$k_4 - k_3$}
		
		\adjmatrixbraceleftI{6}{8}{$d_2$}
		\adjmatrixbraceleftI{5}{5}{$d_3 - d_2$}
		\adjmatrixbraceleftI{3}{4}{$d_4 - d_3$}
		\adjmatrixbraceleftI{1}{2}{$n - d_4$}
	
	\end{tikzpicture}
	\caption{A sorted ordered graph $G$, with $\mathcal{S}_G = \{(d_1, k_1), (d_2, k_2), (d_3, k_3), (d_4, k_4)\}$. \\ Orange blocks indicate the edges of $G$, and green blocks indicate the permitted edges, $\mathcal{P}_G$.}
\end{center}
\end{figure}

\begin{leftbar}
	\begin{lemma}
		\label{lem:permitted_edges}
		Let $G \subseteq K_{n,n}$ be a sorted ordered graph. Then:
		
		\[ a^\star_G = \sum_{\substack{G \subseteq H \in EL_n \\ \left(E(H) \setminus E(G)\right) \subseteq \mathcal{P}_G}} (-1)^{|E(H) \setminus E(G)|} \]
	\end{lemma}
\end{leftbar}
\begin{proof}
	Let $S = E(K_{n,n}) \setminus (\mathcal{P}_G \cupdot E(G))$. By Lemma \ref{lem:dual_coeff_upper_sum}, Corollary \ref{cor:dual_coeff_diff_sum}, and using the inclusion-exclusion principle, we have:
	\begin{equation*}
		\begin{split}
			a^\star_G &= (-1)^{|E(G)|+1} \sum_{\substack{H \supseteq G \\ H \in MC_n}} (-1)^{\chi(H)} \\
			&= \sum_{\substack{G \subseteq H \in EL_n \\ \left(E(H) \setminus E(G)\right) \subseteq \mathcal{P}_G}} (-1)^{|E(H) \setminus E(G)|} + (-1)^{|E(G)|+1} \sum_{\substack{G \subseteq H \in MC_n \\ E(H) \cap S \ne \emptyset}} (-1)^{\chi(H)} \\
		&= \sum_{\substack{G \subseteq H \in EL_n \\ \left(E(H) \setminus E(G)\right) \subseteq \mathcal{P}_G}} (-1)^{|E(H) \setminus E(G)|} + (-1)^{|E(G)|+1} \sum_{\emptyset \ne T \subseteq S} (-1)^{|T|} \sum_{(G \cupdot T) \subseteq H \in MC_n} (-1)^{\chi(H)}
		\end{split}
	\end{equation*}
	
	Observe that for every set $\emptyset \ne T \subseteq S$, the graph $G \cupdot T$ is not totally ordered. Therefore, by Lemma \ref{lem:non_tot_ord_zero_coeff}, every summand $\sum_{(G \cupdot T) \subseteq H \in MC_n} (-1)^{\chi(H)}$ in the above sum is zero, thus concluding the proof.
\end{proof}

\subsubsection{Factorizing into $\langle n,d,k \rangle$-blocks}

Having shown that only ``permitted edges'' need be considered, our next step is to reduce the computation of the dual coefficient $a^\star_G$, to that of dual coefficients of \textit{simpler} graphs. In order to do so, we must first handle the following ``degenerate'' case.

\begin{leftbar}
	\begin{lemma}
		\label{lem:degenerate_sequence}
		Let $G \subseteq K_{n,n}$ be a sorted ordered graph and let $\mathcal{S}_G = \{(d_1, k_1), \dots, (d_t, k_t)\}$ be the representing sequence of $G$. If there exists some $i \in [t-1]$ such that $d_{i+1} \le k_i$, then $a^\star_G = 0$.
	\end{lemma}
\end{leftbar}
\begin{proof}
	Let $i \in [t-1]$ such that $d_{i+1} \le k_i$, and let $X = \{a_1, \dots, a_{k_i}\} \subsetneq \{a_1, \dots, a_n\}$. By Lemma \ref{lem:permitted_edges}:
	
	\[ a^\star_G = \sum_{\substack{G \subseteq H \in EL_n \\ \left(E(H) \setminus E(G)\right) \subseteq \mathcal{P}_G}} (-1)^{|E(H) \setminus E(G)|} \]
	
	Therefore, it suffices to show that any graph $H \supseteq G$ with $\left(E(H) \setminus E(G)\right) \subseteq \mathcal{P}_G$ is not elementary. Let $H$ be such a graph, then $|N_{H}(X)| \le d_{i+1} \le k_i = |X|$ and by Theorem \ref{thm:hetyei_conditions}, $H$ is indeed not elementary. 
\end{proof}

Any sorted ordered graph $G$ whose representing sequence is not degenerate in the above sense, can be neatly factorized into a set of $\langle n,d,k \rangle$-blocks. In the following lemma we construct such a decomposition, and relate the dual coefficients of the each component with that of the original graph.

\begin{leftbar}
	\begin{lemma}
		\label{lem:block_decompose}
		Let $G \subseteq K_{n,n}$ be a sorted ordered graph. Let $\mathcal{S}_G = \{(d_1, k_1), \dots, (d_t, k_t)\}$ be the representing sequence of $G$, where $\forall i \in [t-1]:\ d_{i+1} > k_{i}$. Denote $k_0 = 0$, $d_{t+1} = n$, and:
		\begin{equation*}
			\begin{split}
				\forall i \in [t]:\ &A_i = \{a_{k_{i-1}+1}, \dots, a_{d_{i+1}}\},\quad B_i = \{b_{k_{i-1}+1}, \dots, b_{d_{i+1}}\}
			\end{split}
		\end{equation*}
		Furthermore, for all $i \in [t]$, let $G_i = G\left[A_i \cupdot B_i \right]$ be the induced graph on the vertices $A_i \cupdot B_i$. Then:
		
		\[ a^\star_G = \prod_{i = 1}^{t} a^\star_{G_i} \]
	\end{lemma}
\end{leftbar}
\begin{proof}
	
	For all $i \in [t]$, let $S_i = \{ a_{k_{i-1}+1}, \dots, a_{k_i}\}$ and $T_i = \{b_{d_i + 1}, \dots, b_{d_{i+1}}\}$. Observe that the permitted edges for $G$ are partitioned by the sets $S_i$, $T_i$ as follows: $\mathcal{P}_G = \bigcupdot_{i = 1}^{t} (S_i \times T_i)$. Furthermore, since $\forall i \in [t]:\ d_i > k_{i-1}$, we have:
	
	\[ \forall i \in [t]:\ (A_i \times B_i) \cap \mathcal{P}_G = (S_i \times T_i) \]
	
	Thus, each induced graph $G_i$ ``covers'' the set $(S_i \times T_i)$, and the set of all induced graphs covers all the permitted edges $\mathcal{P}_G$. Since $d_i > k_{i-1}$, then $\forall i \in [t-1]:$ $G_i$ has at least one left vertex, $a_{d_{i+1}}$, whose neighbour set in $G_i$ is the entire right bipartition $B_i$. Similarly, in $G_t$ the neighbour set of the right vertex $b_{k_{t-1}+1}$ is the entire left bipartition $A_t$. Thus, by Corollary \ref{cor:dual_coeff_diff_sum}:
	
	\[ \forall i \in [t]:\ a^\star_{G_i} = \sum_{\substack{G_i \subseteq H \\ H \text{ is elementary}}} (-1)^{|E(H) \setminus E(G)|}\]
	
	To complete the proof, it therefore remains to show a bijection between elementary completions of $G$ using the permitted edges $\mathcal{P}_G$, and elementary completions of each of the graphs $G_i$.
	
	\begin{figure}[h!]
\begin{center}
	\begin{tikzpicture}[baseline=0cm,adjmatrixenv]
		\matrix [adjmatrix,outer ysep=0.7pt,inner sep=8pt, row sep=1em, column sep=1.4em] (m)  
		{
			\pad & \pad & \pad & \pad & \pad & \pad & \pad & \pad  \\
			\pad & \pad & \pad & \pad & \pad & \pad & 1    & 1     \\
			\pad & \pad & \pad & \pad & \pad & \pad & 1    & 1     \\
			\pad & \pad & \pad & \pad & 1    & 1    & 1    & 1     \\
			\pad & \pad & \pad & \pad & 1    & 1    & 1    & 1     \\
			\pad & \pad & 1    & 1    & 1    & 1    & 1    & 1     \\
			\pad & \pad & 1    & 1    & 1    & 1    & 1    & 1     \\
			\pad & \pad & 1    & 1    & 1    & 1    & 1    & 1     \\
		};
		
		\node [fit= \foreach \X in {6,...,8}{(m-\X-1)}
					\foreach \X in {6,...,8}{(m-\X-2)}]
		[draw=green, fill=green, fill opacity=0.1, thick,inner sep=2.6pt] (per-rect1) {};
	
		\node [fit= \foreach \X in {4,...,5}{(m-\X-3)}
					\foreach \X in {4,...,5}{(m-\X-4)}]
		[draw=green, fill=green, fill opacity=0.1, thick,inner sep=2.6pt] (per-rect2) {};
	
		\node [fit= \foreach \X in {2,...,3}{(m-\X-5)}
					\foreach \X in {2,...,3}{(m-\X-6)}]
		[draw=green, fill=green, fill opacity=0.1, thick,inner sep=2.6pt] (per-rect3) {};
		
		\node [fit= \foreach \X in {1,...,1}{(m-\X-7)}
					\foreach \X in {1,...,1}{(m-\X-8)}]
		[draw=green, fill=green, fill opacity=0.1, thick,inner sep=2.6pt] (per-rect4) {};
			
		\node [fit= \foreach \X in {6,...,8}{(m-\X-3)}
					\foreach \X in {6,...,8}{(m-\X-4)}]
		[draw=niceorange, thick,inner sep=2.6pt] (block-rect1) {};
	
		\node [fit= \foreach \X in {4,...,8}{(m-\X-5)}
					\foreach \X in {4,...,8}{(m-\X-6)}]
		[draw=niceorange, thick,inner sep=2.6pt] (block-rect2) {};
	
		\node [fit= \foreach \X in {2,...,8}{(m-\X-7)}
					\foreach \X in {2,...,8}{(m-\X-8)}]
		[draw=niceorange, thick,inner sep=2.6pt] (block-rect3) {};
	
		\node [fit= \foreach \X in {6,...,8}{(m-\X-1)}
					\foreach \X in {6,...,8}{(m-\X-2)}
					\foreach \X in {6,...,8}{(m-\X-3)}]
		[draw=blue, fill=blue, draw opacity=0.3, fill opacity=0.1, thick,inner sep=2.6pt] (block-g1) {};
	
		\node [fit= \foreach \X in {4,...,6}{(m-\X-3)}
					\foreach \X in {4,...,6}{(m-\X-4)}
					\foreach \X in {4,...,6}{(m-\X-5)}]
		[draw=blue, fill=blue, draw opacity=0.3, fill opacity=0.1, thick,inner sep=2.6pt] (block-g2) {};
	
		\node [fit= \foreach \X in {2,...,4}{(m-\X-5)}
					\foreach \X in {2,...,4}{(m-\X-6)}
					\foreach \X in {2,...,4}{(m-\X-7)}]
		[draw=blue, fill=blue, draw opacity=0.3, fill opacity=0.1, thick,inner sep=2.6pt] (block-g3) {};
		
		\node [fit= \foreach \X in {1,...,2}{(m-\X-7)}
					\foreach \X in {1,...,2}{(m-\X-8)}]
		[draw=blue, fill=blue, draw opacity=0.3, fill opacity=0.1, thick,inner sep=2.6pt] (block-g4) {};
		
		\adjmatrixbracebottom{2}{1}{$k_1$}
		\adjmatrixbracebottom{4}{3}{$k_2 - k_1$}
		\adjmatrixbracebottom{6}{5}{$k_3 - k_2$}
		\adjmatrixbracebottom{8}{7}{$k_4 - k_3$}
		
		\adjmatrixbraceleftI{6}{8}{$d_2$}
		\adjmatrixbraceleftI{4}{5}{$d_3 - d_2$}
		\adjmatrixbraceleftI{2}{3}{$d_4 - d_3$}
		\adjmatrixbraceleftI{1}{1}{$n - d_4$}
	
	\end{tikzpicture}
	\caption{A sorted ordered graph $G$, with $\mathcal{S}_G = \{(d_1, k_1), (d_2, k_2), (d_3, k_3), (d_4, k_4)\}$. The permitted edges $\mathcal{P}_G$ are covered. Orange blocks indicate the edges of $G$, green blocks indicate the permitted edges, and blue blocks are the induced graphs $G_i$}
\end{center}
\end{figure}
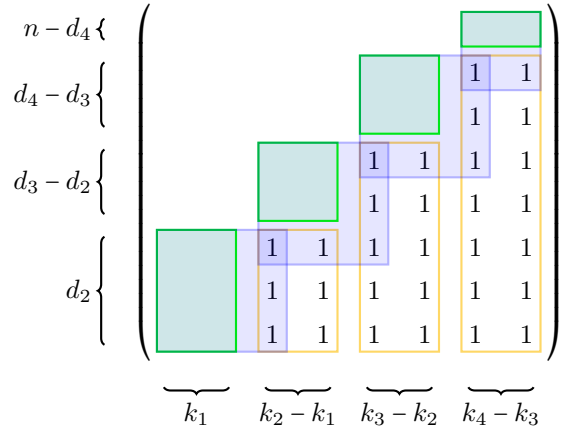
	
	Let $G \subseteq H \in EL_n$ such that $\left(E(H) \setminus E(G)\right) \subseteq \mathcal{P}_G$. For all $i \in [t]$, let $H_i = H\left[ A_i \cupdot B_i \right]$. Since $H \supseteq G$, clearly also $\forall i \in [t]:\ H_i \supseteq G_i$. It remains to show that every such $H_i$ is elementary. To this end, we use Theorem \ref{thm:hetyei_conditions}: let $i \in [t]$ and let $\emptyset \ne X \subsetneq A_i$. If $X \cap A_{i+1} \ne \emptyset$ then $H_i$ has a vertex of full degree, and so $|N_{H_i}(X)| = |B_i| = |A_i| > |X|$. Otherwise, if $X \cap A_{i+1} = \emptyset$ then let $X' = X \cupdot \{a_1, \dots, a_{k_{i-1}}\}$. Observe that $N_H(X') = N_H(X) = N_{H_i}(X) \cupdot \{b_1, \dots, b_{k_{i-1}}\}$. However, $H$ is elementary, therefore $|N_{H}(X')| > |X'| = |X| + k_{i-1}$. In both cases we have $|N_{H_i}(X)| > |X|$ and $H_i$ is elementary. 
	
	Conversely, let $H_1 \supseteq G_1, \dots, H_t \supseteq G_t$ be elementary graphs. Then it suffices to show that $H \supseteq G$ whose edges are $E(H) = E(G) \cup E(H_1) \cup E(H_2) \cup \dots \dots E(H_t)$ is also elementary. Let $X \subsetneq A$, let $i$ be the largest index such that $a_i \in X$, and let $j$ be the index for which $k_{j-1} < i \le k_j$. Thus: 
	
	\[N_{H}(X) = N_{H_j}(X \cap A_j) \cupdot \{b_1, \dots, b_{k_{i-1}}\}\]
	
	If $X \cap A_j = A_j$, then $N_{H_j}(X \cap A_j) = B_j$ and thus $|N_{H}(X)| = k_{i-1} + |B_j| = d_{j+1} > k_j \ge |X|$, and indeed $H$ is elementary. Otherwise, since $H_j$ is elementary and $(X \cap A_j) \subsetneq A_j$, we have $|N_{H_j}(X \cap A_j)| > |X \cap A_j|$, and therefore:
	
	\[ |N_{H}(X)| = |N_{H_j}(X \cap A_j)| + k_{i-1} > |X \cap A_j| + k_{i-1} \ge |X| \qedhere \]
\end{proof}

\subsubsection{The Dual Coefficients of $\langle n,d,k \rangle$-blocks}

Finally, having reduced the computation of the dual coefficient of an arbitrary ordered graph $G$ to that of simple ``blocks'', we are left with the task of directly computing the dual coefficient for any such block. 

\begin{leftbar}
	\begin{lemma}
		\label{lem:block_coeff}
		Let $n,d,k \in \mathbb{N}$, such that $0 \le d \le n$, $0 < k < n$. The coefficient of the $\langle n, d, k \rangle$-block is:
		
		\[ a^\star_{\langle n, d, k \rangle} = \begin{dcases*}
		{{n-1} \choose {k}}, & $d = 0$ \\
		-{{n-d-1} \choose {k-d}}
		{{k-1} \choose {d-1}}, & $d > 0$ 
		\end{dcases*}\quad\quad(\star)\]
		
	\end{lemma}
\end{leftbar}
\begin{proof}
	The proof is by induction on $n$, $d$ and $k$. For the base case, let $G$ be an $\langle 2, d, 1 \rangle$-block, where $d \in \{0,1,2\}$. In all three cases, only $K_{2,2} \supseteq G$ is elementary, thus by Corollary \ref{cor:dual_coeff_diff_sum} they all satisfy equation $(\star)$, as required. Next, we use complete induction. Let $G$ be an $\langle n,d,k \rangle$-block, where $n > 2$, and assume equation $(\star)$ holds for all $\langle n', d', k' \rangle$-blocks, such that:
	\[(n' < n) \lor (n' = n\ \land\ k' = k\ \land\ d' > d) \] 
	
	If $d > k$, then $\forall X \subsetneq \{a_1, \dots, a_n\}:\ |N(X)| > |X|$, thus by  Theorem \ref{thm:hetyei_conditions}, $G$ is elementary and by Lemma \ref{lem:mc_dual_coeff_zero}, $a^\star_G = 0$. Otherwise, $d \le k$. In this case, denote $S = \{(a_1, b_n), \dots, (a_k, b_n)\}$, and partition the set of all elementary graphs containing $G$ into two disjoint sets:
	\begin{equation*}
	\begin{split}
	\mathcal{H}_1 &= \set{G \subseteq H \in EL_n}{E(H) \cap S \ne \emptyset} \\
	\mathcal{H}_2 &= \set{G \subseteq H \in EL_n}{E(H) \cap S = \emptyset} \\
	\end{split}
	\end{equation*}	
	
	By Corollary \ref{cor:dual_coeff_diff_sum}, the dual coefficient of $G$ is given by the sum over the aforementioned sets:
	
	\[ a^\star_G = \sum_{H \in \mathcal{H}_1} (-1)^{|E(H) \setminus E(G)|} + \sum_{H \in \mathcal{H}_2} (-1)^{|E(H) \setminus E(G)|}\]
	
	\noindent\underline{The contributions of $\mathcal{H}_1$:} \\
	
	To sum the contributions of all graphs in $\mathcal{H}_1$, we use the inclusion-exclusion principle. First, note that $\BPMnstar$ is invariant to permutations over each bipartition (that is, if $H \cong G$ then $a^\star_G$ = $a^\star_H$). Therefore, for every subset $T \subseteq S$ of selected edges, we may, without loss of generality, ``sort'' the graph to obtain an isomorphic sorted ordered graph. Consequently, denote $\forall t \in [k]:\ G_t = G \cup \{(a_{k-t+1}, b_{d+1}), \dots, (a_{k}, b_{d+1})\}$. By the inclusion-exclusion principle, we have:
	
	\begin{equation*}
	\begin{split}
	\sum_{H \in \mathcal{H}_1} (-1)^{|E(H) \setminus E(G)|} &= \sum_{t=1}^{k} (-1)^{t+1} {k \choose t} \cdot (-1)^t \cdot a^\star_{G_t} = -\sum_{t=1}^{k} {k \choose t} \cdot a^\star_{G_t}
	\end{split}
	\end{equation*}
	
	If $t = k$, then $G_t$ is an $\langle n, d+1, k \rangle$-block, for which the induction hypothesis holds. Otherwise, for $t \in [k-1]$, the biadjacency matrix of each graph $G_t$ can be partitioned into blocks, as follows:
	
	\begin{center}
\begin{tikzpicture}[baseline=0cm,adjmatrixenv]
	\matrix [adjmatrix,outer ysep=0.7pt,inner sep=8pt, row sep=1em, column sep=1.4em] (m)  
	{
		\phantom{0} &   &   &   &   &   & 1 & 1  \\
		\phantom{0} &   &   &   &   &   & 1 & 1  \\
		\phantom{0} &   &   &   &   &   & 1 & 1  \\
		\phantom{0} &   &   &   &   &   & 1 & 1  \\
		\phantom{0} &   &   & 1 & 1 & 1 & 1 & 1  \\
		1           & 1 & 1 & 1 & 1 & 1 & 1 & 1  \\
		1           & 1 & 1 & 1 & 1 & 1 & 1 & 1  \\
		1           & 1 & 1 & 1 & 1 & 1 & 1 & 1  \\
	};
	
	\node [fit= \foreach \X in {5,...,8}{(m-\X-1)}
				\foreach \X in {5,...,8}{(m-\X-2)}
				\foreach \X in {5,...,8}{(m-\X-3)}
				\foreach \X in {5,...,8}{(m-\X-4)}]
	[draw=blue, fill=blue, fill opacity=0.1, thick,inner sep=2.6pt] (cut-rect1) {};

	\node [fit= \foreach \X in {2,...,6}{(m-\X-4)}
				\foreach \X in {2,...,6}{(m-\X-5)}
				\foreach \X in {2,...,6}{(m-\X-6)}
				\foreach \X in {2,...,6}{(m-\X-7)}
				\foreach \X in {2,...,6}{(m-\X-8)}]
	[draw=green, fill=green, fill opacity=0.1, thick,inner sep=2.6pt] (cut-rect2) {};

		\node [fit= \foreach \X in {1,...,2}{(m-\X-7)}
					\foreach \X in {1,...,2}{(m-\X-8)}]
	[draw=niceorange, fill=niceorange, fill opacity=0.1, thick,inner sep=2.6pt] (cut-rect3) {};
		
	\adjmatrixbracebottom{3}{1}{$k-t$}
	\adjmatrixbracebottom{6}{4}{$t$}
	\adjmatrixbracebottom{8}{7}{$n-k$}
	
	\adjmatrixbraceleftI{6}{8}{$d$}
	\adjmatrixbraceleftI{5}{5}{$1$}
	\adjmatrixbraceleftI{1}{4}{$n-d-1$}	

\end{tikzpicture}
\end{center}
	
	If $t < k-d$, then by Definition \ref{def:permitted_edges} the permitted edges for the vertices $\{a_1, \dots, a_{k-t}\}$ in $G_t$ are only those connecting them to $b_{d+1}$. Therefore, $G_t$ cannot be completed to an elementary graph using only permitted edges, and by Lemma \ref{lem:permitted_edges}, $a^\star_{G_t} = 0$. Otherwise, by Lemma \ref{lem:block_decompose}, the coefficient $a^\star_{G_t}$ is the product of coefficients for each of the three blocks. The first two are an $\langle d+1, d, k-t \rangle$-block and a $\langle n-k+t, d+1-k+t, t \rangle$-block. The third is a complete bipartite graph over $n-k$ vertices, and thus does not affect the coefficient of $G$.
	
	\begin{figure}[h]  
\centering 
\begin{subfigure}[b]{0.4\linewidth}
	\begin{tikzpicture}[baseline=0cm,adjmatrixenv]
		\matrix [adjmatrix,outer ysep=0.7pt,inner sep=8pt, row sep=1em, column sep=1.4em] (m)  
		{
			\phantom{0} &   &   & 1 & 1  \\
			1           & 1 & 1 & 1 & 1  \\
			1           & 1 & 1 & 1 & 1  \\
			1           & 1 & 1 & 1 & 1  \\
			1           & 1 & 1 & 1 & 1  \\
		};
			
		\adjmatrixbracebottom{3}{1}{$k-t$}
		\adjmatrixbraceright{1}{5}{$d+1$}
		\adjmatrixbraceleftI{2}{5}{$d$}
	
		\node [fit= \foreach \X in {2,...,5}{(m-\X-1)}
					\foreach \X in {2,...,5}{(m-\X-2)}
					\foreach \X in {2,...,5}{(m-\X-3)}]
		[draw=blue, thick,inner sep=2.6pt] (rect1) {};
	
		\node [fit= \foreach \X in {1,...,5}{(m-\X-4)}
					\foreach \X in {1,...,5}{(m-\X-5)}]
					[draw=blue, thick,inner sep=2.6pt] (rect2) {};
	\end{tikzpicture}
	\caption{$\langle d+1, d, k-t \rangle$-block}
\end{subfigure}
\begin{subfigure}[b]{0.4\linewidth}
	\begin{tikzpicture}[baseline=0cm,adjmatrixenv]
	\matrix [adjmatrix,outer ysep=0.7pt,inner sep=8pt, row sep=1em, column sep=1.4em] (m)  
	{
		\phantom{0} &   &   & 1 & 1  \\
		\phantom{0} &   &   & 1 & 1  \\
		\phantom{0} &   &   & 1 & 1  \\
		1           & 1 & 1 & 1 & 1  \\
		1           & 1 & 1 & 1 & 1  \\
	};
	
	\adjmatrixbracebottom{3}{1}{$t$}
	\adjmatrixbraceright{1}{5}{$n - k + t$}
	\adjmatrixbraceleftI{4}{5}{$d+1-k+t$}

	\node [fit= \foreach \X in {4,...,5}{(m-\X-1)}
				\foreach \X in {4,...,5}{(m-\X-2)}
				\foreach \X in {4,...,5}{(m-\X-3)}]
	[draw=green, thick,inner sep=2.6pt] (rect1) {};
	
	\node [fit= \foreach \X in {1,...,5}{(m-\X-4)}
				\foreach \X in {1,...,5}{(m-\X-5)}]
	[draw=green, thick,inner sep=2.6pt] (rect2) {};
	
	\end{tikzpicture}
	\caption{$\langle n-k+t, d+1-k+t, t \rangle$-block}
\end{subfigure}
\end{figure}
	
	Observe that if $k-d < t < k$, then the $\langle d+1, d, k+1 \rangle$-block is elementary, thus by Lemma \ref{lem:mc_dual_coeff_zero}, its dual coefficient is zero. Consequently, only two potentially non-zero cases remain: $t=k$ and $t=k-d$. For both cases, the induction hypothesis holds. Observe that if $d = 0$, both cases converge to a single case. Thus: \\
	
	\underline{$d > 0$:}
	\begin{equation*}
	\begin{split}
	-\sum_{t=1}^{k} {k \choose t} \cdot a^\star_{G_t} &=
	- {k \choose {k-d}} \cdot a^\star_{\langle d+1, d, d \rangle} - {k \choose k} \cdot a^\star_{\langle n, d+1, k \rangle} \\
	&= - {{n - d - 2} \choose {k-d-1}} \cdot \left[ {{k} \choose {k-d}} + {{k-1} \choose d}\right] = - {{n - d - 2} \choose {k-d-1}} {{k-1} \choose {d-1}}
	\end{split}
	\end{equation*}
	
	\underline{$d = 0$:}
	\begin{equation*}
	\begin{split}
	-\sum_{t=1}^{k} {k \choose t} \cdot a^\star_{G_t} &= 
	- {k \choose k} \cdot a^\star_{\langle n, 1, k \rangle} = {{n - 2} \choose {k-1}} 
	\end{split}
	\end{equation*}
	
	\noindent\underline{The contributions of $\mathcal{H}_2$:} \\
	
	If $k = n-1$, then $\mathcal{H}_2 = \emptyset$, thus there are no contributions from $\mathcal{H}_2$. This assertion follows since for any $H \supseteq G$ with $E(H) \cap S = \emptyset$, we have $|N(\{a_1, \dots, a_{n-1}\})| \le n-1 = |\{a_1, \dots, a_{n-1}\}|$. Thus, by Theorem \ref{thm:hetyei_conditions}, $H$ is not elementary. Otherwise, if $k < n-1$, we claim that: 
	
	\[ \sum_{H \in \mathcal{H}_2} (-1)^{|E(H) \setminus E(G)|} = a^\star_{\langle n-1, d, k \rangle} \]
	
	Since the induction hypothesis holds for the $\langle n-1, d, k \rangle$-block, proving the above identity would yield an expression for the contributions of $\mathcal{H}_2$. Denote the $\langle n-1, d, k \rangle$-block by $G'$. To prove the aforementioned identity, it remains to show a bijection between elementary graphs $G' \subseteq H' \in EL_{n-1}$, and elementary graphs $G \subseteq H \in EL_n$, where $E(H) \cap S = \emptyset$. Furthermore, we must also maintain $|E(H') \setminus E(G')| = |E(H) \setminus E(G)|$, for any two graphs $H'$ and $H$ which are mapped to one another by the bijection. The bijection is defined as follows: \\
	
	\underline{$H \mapsto H'$}: \\
	
	Let $H \supseteq G$ be an elementary graph such that $E(H) \cap S = \emptyset$. We claim that $H' = H - a_n - b_n$ is also elementary. By Theorem \ref{thm:hetyei_conditions}, it suffices to show that $\forall X \subsetneq \{a_1, \dots, a_{n-1}\}:\ |N_{H'}(X)| > |X|$. If $X \cap \{a_{k+1}, \dots, a_{n-1}\} \ne \emptyset$, then $N_{H'}(X) = \{b_1, \dots, b_{n-1}\}$. Thus $|N_{H'}(X)| = n-1 > |X|$. Otherwise, if $X \cap \{a_{k+1}, \dots, a_{n-1}\} = \emptyset$, then $N_{H'}(X) = N_{H}(X)$ and therefore: $|N_{H'}(X)| = |N_{H}(X)| > |X|$, as required. \\
	
	\underline{$H' \mapsto H$}: \\
	
	Let $H' \supseteq G'$ be an elementary graph. We claim that the graph $H \subseteq K_{n,n}$ where $E(H) = E(H') \cupdot \{(a_{k+1}, b_n), \dots, (a_n, b_n)\}$, is also elementary. Once again, we use Theorem \ref{thm:hetyei_conditions}. Let $X \subsetneq \{a_1, \dots, a_n\}$. If $X \cap \{a_{k+1}, \dots, a_{n}\} \ne \emptyset$, then $N_{H}(X) = \{b_1, \dots, b_{n}\}$. Thus $|N_{H}(X)| = n > |X|$. Otherwise, if $X \cap \{a_{k+1}, \dots, a_{n}\} = \emptyset$, then $N_{H}(X) = N_{H'}(X)$ and therefore: $|N_{H}(X)| = |N_{H'}(X)| > |X|$, as required. \\
	
	\noindent\underline{Summing up the contributions:} \\
	
	Finally, we have reduced the computation of the coefficient $a^\star_{\langle n,d,k \rangle}$ to a sum of coefficients $a^\star_{\langle n', d', k' \rangle}$ for which the induction hypothesis holds. It now remains to sum up the contributions for each possible case. If $d > 0$ and $k = n-1$, then: 
	\begin{equation*}
	\begin{split}
	a^\star_G &= \sum_{H \in \mathcal{H}_1} (-1)^{|E(H) \setminus E(G)|} = - {{n - d - 2} \choose {k-d-1}} {{k-1} \choose {d-1}} = - {{n - d - 1} \choose {k-d}} {{k-1} \choose {d-1}}
	\end{split}
	\end{equation*}
	
	If $d > 0$ and $k < n-1$, then:
	\begin{equation*}
	\begin{split}
	a^\star_G &= \sum_{H \in \mathcal{H}_1} (-1)^{|E(H) \setminus E(G)|} + \sum_{H \in \mathcal{H}_2} (-1)^{|E(H) \setminus E(G)|} \\
	&= - {{n - d - 2} \choose {k-d-1}} {{k-1} \choose {d-1}} - {{n - d - 2} \choose {k-d}} {{k-1} \choose {d-1}} = - {{n - d - 1} \choose {k-d}} {{k-1} \choose {d-1}}
	\end{split}
	\end{equation*}
	
	If $d = 0$ and $k = n-1$, then:
	\begin{equation*}
	\begin{split}
	a^\star_G &= \sum_{H \in \mathcal{H}_1} (-1)^{|E(H) \setminus E(G)|} = {{n - 2} \choose {k-1}} = {{n - 1} \choose {k}}
	\end{split}
	\end{equation*}
	
	Lastly, if $d = 0$ and $k < n-1$, then:
	\begin{equation*}
	\begin{split}
	a^\star_G &= \sum_{H \in \mathcal{H}_1} (-1)^{|E(H) \setminus E(G)|} + \sum_{H \in \mathcal{H}_2} (-1)^{|E(H) \setminus E(G)|} \\
	&= {{n - 2} \choose {k-1}} + {{n - 2} \choose {k}} = {{n - 1} \choose {k}} \qedhere
	\end{split}
	\end{equation*}
\end{proof}

\subsubsection{Putting It Together}

\begin{leftbar}
\begin{theorem}[Theorem \ref{thm:bpm_star_poly}, restated]
	
	Let $G \subseteq K_{n,n}$. If $G$ is not totally ordered, then $a^\star_G = 0$. Otherwise, let $\mathcal{S}_G = \{(d_1, k_1), \dots, (d_t, k_t)\}$ be the representing sequence of $G$, and let $k_0=0$. Then:
	
	\begin{equation*}
		\begin{split}
		a^\star_G = {{n-k_{t-1}-1} \choose {n - d_t}} \cdot \prod_{i=1}^{t-1} f\left(d_{i+1} - k_{i-1}, d_i - k_{i-1}, k_i - k_{i-1}\right)
		\end{split}
	\end{equation*}
	
	where:
	\begin{equation*}
		\begin{split}
		f(n,d,k) = \begin{dcases*}
		{{n-1} \choose {k}}, & $d \le 0$ \\
		- {{n-d-1} \choose {k-d}} {{k-1} \choose {d-1}}, & $d>0$
		\end{dcases*}
		\end{split}
	\end{equation*}
\end{theorem}
\end{leftbar}
\begin{proof}
	Let $G \subseteq K_{n,n}$. If $G$ is not totally ordered, then by Lemma \ref{lem:non_tot_ord_zero_coeff}, $a^\star_G = 0$. Otherwise, if $G$ is totally ordered and there exists some $i \in [t-1]$ such that $d_{i+1} \le k_i$, then by Lemma \ref{lem:degenerate_sequence}, $a^\star_G = 0$, and indeed:
	
	\[ f\left(d_{i+1} - k_{i-1}, d_i - k_{i-1}, k_i - k_{i-1}\right) = {{d_{i+1} - k_{i-1} - 1} \choose {k_i - k_{i-1}}} = 0 \]
	
	Finally, if $G$ is totally ordered, and $\forall i \in [t-1]:\ d_{i+1} > k_i$, then let 
	$\forall i \in [t]:\ A_i = \{a_{k_{i-1}+1}, \dots, a_{d_{i+1}}\}$, $B_i = \{b_{k_{i-1}+1}, \dots, b_{d_{i+1}}\}$, where $k_0 = 0$ and $d_{t+1} = n$. Furthermore $\forall i \in [t]$, let $G_i = G\left[A_i \cupdot B_i \right]$ be the induced graph on the vertices $A_i \cupdot B_i$. By Lemma \ref{lem:block_decompose}:
	
	\[ a^\star_G = \prod_{i = 1}^{t} a^\star_{G_i} \] 
	
	Observe that $\forall i \in [t-1]$, the graph $G_i$ is an $\langle d_{i+1} - k_{i-1}, d_i - k_{i-1}, k_i - k_{i-1} \rangle$-block. Thus, its coefficient is given by the expression in Lemma \ref{lem:block_coeff}. However, the last graph $G_t$ may \textit{not} be a ``block''. In fact, there are two possible cases: either $d_t = n$, in which case $G_t$ is a complete bipartite graph, and thus $a^\star_G = 1$. Otherwise, $d_t < n$, and $G_t$ is a biclique joining $n - k_{t-1}$ left vertices to $n - d_{t}$ right vertices. In this case, since $a^\star_G$ is invariant to swapping the two bipartitions, then without loss of generality we may do so, thus obtaining an isomorphic $\langle n - k_{t-1}, 0, n - d_{t} \rangle$-block. Thus, we have:
	
	\[ a^\star_{G_t} = \begin{dcases*}
		1, &$d_t = n$ \\
		{{n-k_{t-1}-1} \choose {n - d_t}}, & $d_t < n$
	\end{dcases*}\  =\  {{n-k_{t-1}-1} \choose {n - d_t}} \]
	
	Putting it all together, we obtain:
	\begin{equation*}
		\begin{split}
				 a^\star_G &= \prod_{i = 1}^{t} a^\star_{G_i} \\
				 &= {{n-k_{t-1}-1} \choose {n - d_t}} \cdot \prod_{i = 1}^{t-1} a^\star_{\langle d_{i+1} - k_{i-1}, d_i - k_{i-1}, k_i - k_{i-1}\rangle} \\
				 &= {{n-k_{t-1}-1} \choose {n - d_t}} \cdot \prod_{i=1}^{t-1} f\left(d_{i+1} - k_{i-1}, d_i - k_{i-1}, k_i - k_{i-1}\right) \qedhere
		\end{split}
	\end{equation*}
\end{proof}

\subsection{Corollaries of Theorem \ref{thm:bpm_star_poly}: The $\ell_1$-norms of $\BPMn$}

Theorem \ref{thm:bpm_star_poly} allows us to compute the dual coefficient of any graph $G \subseteq K_{n,n}$. It is not hard to see that for some graphs $G \subseteq K_{n,n}$, the coefficient $a^\star_G$ may be \textit{exponential} in $n$. For instance, the biclique $K_{n, \sfrac{n}{2}}$ is an ordered graph whose representing sequence is $\left(\frac{n}{2}, n\right)$. Therefore:

\[a^\star_{K_{n, \sfrac{n}{2}}} = {{n-1} \choose {\frac{n}{2}}} \sim \frac{2^n}{\poly\left(n\right)} \]

The aforementioned bound is qualitatively tight. Namely, for \textit{every} graph $G \subseteq K_{n,n}$, the coefficient $a^\star_G$ is \textit{at most} exponential in $2n$:

\begin{leftbar}
	\begin{lemma}
		\label{lem:bound_dual_coeff}
		Let $G \subseteq K_{n,n}$. The dual coefficient of $G$ is bounded by $|a^\star_G| \le 2^{2n}$. 
	\end{lemma}
\end{leftbar}
\begin{proof}
	If $G$ is not totally ordered then by Lemma \ref{lem:non_tot_ord_zero_coeff} $a^\star_G = 0$. Otherwise, let $\{(d_1, k_1), \dots, (d_t, k_t)\}$ be the \textit{representing sequence} of $G$, let $k_0 = 0$, and let:
	\begin{equation*}
	\begin{split}
	f(n,d,k) = \begin{dcases*}
	{{n-1} \choose {k}}, & $d \le 0$ \\
	- {{n-d-1} \choose {k-d}} {{k-1} \choose {d-1}}, & $d>0$
	\end{dcases*}
	\end{split}
	\end{equation*}
	
	Observe that both in the case $d \le 0$ and in the case $d>0$, we have $|f(n,d,k)| \le 2^{n-d+k}$ (bounding each binomial coefficient using the binomial theorem). Therefore, using the expression given by Theorem \ref{thm:bpm_star_poly}:
	
	\begin{equation*}
	\begin{split}
	|a^\star_G| &= {{n-k_{t-1}-1} \choose {n - d_t}} \cdot \prod_{i=1}^{t-1} |f\left(d_{i+1} - k_{i-1}, d_i - k_{i-1}, k_i - k_{i-1}\right)| \\
	&\le 2^{n-k_{t-1}} \cdot \prod_{i=1}^{t-1} 2^{d_{i+1} - k_{i-1} - (d_i - k_{i-1}) +  k_i - k_{i-1}} \\
	&\le 2^{n-k_{t-1}} \cdot 2^{d_{t} + k_{t-1}} \\
	&= 2^{n + d_{t}} \le 2^{2n} \qedhere
	\end{split}
	\end{equation*}
	
\end{proof}

Thus, the multilinear polynomial representing $\BPMnstar$ is ``simple'' in the following sense: its $\ell_1$-norm (the sum of absolute values of its coefficients) is small, over both the $\{0,1\}$ and Fourier basis. 

\begin{leftbar}
	\begin{corollary}
		\label{cor:bpm_n_star_low_l1_norm}
		Let $n > 2$. Then:
		\[ \norm{\BPMnstar}_1 = 2^{\Theta(n \log n)},\ \text{and furthermore}\  \fouriernorm{\BPMn}_1 = \fouriernorm{\BPMnstar}_1 = 2^{\mathcal{O}(n \log n)}\]
	\end{corollary}
\end{leftbar}
\begin{proof}
	In \cite{beniamini2020bipartite}, the number of monomials in $\BPMnstar$ was bounded by:
	\[ (n!)^2 \le \left|\mon\left(\BPMnstar\right)\right| \le (n+2)^{2n + 2} \]
	thus, using Lemma \ref{lem:bound_dual_coeff}, we deduce:
	
	\[ (n!)^2 \le \norm{\BPMnstar}_1 \le (n+2)^{2n + 2} \cdot 2^{2n} \ \Rightarrow\   \norm{\BPMnstar}_1 = 2^{\Theta(n \log n)} \]
	
	As for the Fourier $\ell_1$-norm, we note that the magnitudes of the Fourier coefficients of any Boolean function and its dual are identical (see e.g., \cite{o2014analysis}), therefore $\fouriernorm{\BPMn}_1 = \fouriernorm{\BPMnstar}_1$. Furthermore, recall that $\forall S \subseteq [n]$, $\fouriernorm{\ANDS} = \fouriernorm{\Pi_{i \in S} x_i} = 1$. Thus, by subadditivity and homogeneity:
	
	\begin{equation*}
	\begin{split}
		\fouriernorm{\BPMn}_1 &= \fouriernorm{\BPMnstar}_1 \le \sum_{G \subseteq K_{n,n}} |a^\star_G| \cdot \fouriernorm{\ANDG} = \norm{\BPMnstar}_1 = 2^{\mathcal{O}(n \log n)} \qedhere
	\end{split}
	\end{equation*}
\end{proof}
\section{The Upper Bound on $\epsadeg{\BPMn}$}
\label{section:upper_bound}

In this section we obtain an upper bound on the approximate degree of the bipartite perfect matching function, which holds even for exponentially small values of $\epsilon$.

\begin{leftbar}
	\begin{theorem}
		\label{thm:upper_bound_apx_deg}
		Let $n > 1$ and let $2^{-n \log n} \le \epsilon \le \tfrac{1}{3}$. The $\epsilon$-approximate degree of $\BPMn$ is bounded by:
		
		\[ \epsadeg{\BPMn} = \mathcal{O}(n^{\sfrac{3}{2}} \sqrt{\log n}) \]
	\end{theorem}
\end{leftbar}

This bound is essentially a corollary Theorem \ref{thm:bpm_star_poly}, alongside two further observations. First, we show that the approximate degree of any Boolean function and its dual are identical (for all $\epsilon > 0$). We then prove that Boolean functions whose representing polynomials have small $\ell_1$-norm over the $\{0,1\}$ basis, can be efficiently approximated by low degree polynomials. The latter approach was also employed by Sherstov in \cite{sherstov2020algorithmic}. Let us remark that, to obtain the upper bound on the approximate degree of $\BPMn$ it would have sufficed to merely show that the magnitudes of all dual coefficients are, at most, exponential in $\Theta(n \log n)$. However, we do not know of a simpler proof of this fact, other than leveraging the complete characterization of $\BPMnstar$ given by Theorem \ref{thm:bpm_star_poly}.

\begin{leftbar}
	\begin{lemma}
		\label{lemma:apx_deg_dual}
		Let $f: \{0,1\}^n \rightarrow \{0,1\}$ be a Boolean function, and let $f^\star$ be its Boolean dual. Then:
		\[ \forall 0 < \epsilon < \frac{1}{2}:\ \  \epsadeg{f} = \epsadeg{f^\star} \]
	\end{lemma}
\end{leftbar}
\begin{proof}
	Let $\epsilon > 0$, and let $p \in \mathbb{R}\left[x_1, \dots, x_n\right]$ be a real polynomial that $\epsilon$-approximates $f$ pointwise. Let $p^\star \in \mathbb{R}\left[x_1, \dots, x_n\right]$ be the real polynomial defined by: $p^\star(x_1, \dots, x_n) = 1 - p(1 - x_1, \dots, 1 - x_n)$ (i.e., replace each variable $x_i$ with $(1 - x_i)$, negate all coefficients, and add $1$). Observe that $\deg(p^\star) \le \deg(p)$, since $p^\star$ is obtained by a linear transformation on $p$, thus the degree cannot increase. Furthermore, $\forall x_1, \dots, x_n \in \{0,1\}$, we have: 
	
	\begin{equation*}
	\begin{split}
	|f^\star(x_1, \dots, x_n) - p^\star(x_1, \dots, x_n)| &= \left| 1 - f(1-x_1, \dots, 1-x_n) - (1 - p(1 - x_1, \dots, 1 - x_n))\right| \\
	&= \left|f(1-x_1, \dots, 1-x_n) - p(1-x_1, \dots, 1-x_n)\right| \le \epsilon
	\end{split}
	\end{equation*}
	
	The converse similarly follows, since $(f^\star)^\star = f$.
\end{proof}

For the second lemma, we require a well known Theorem regarding the approximate degree of the $\ANDn$ function. Nisan and Szegedy \cite{nisan1994degree} first showed that for the regime of $\epsilon = \Theta(1)$, we have $\thirdadeg{\ANDn} = \Theta(\sqrt{n})$. Their result was extended by Buhrman, Cleve, De Wolf and Zalka \cite{buhrman1999bounds}, who determined the approximate degree of $\AND$ for any $\epsilon > 0$.

\begin{leftbar}
	\begin{theorem}[\cite{buhrman1999bounds}]
		\label{thm:epsilon_apx_deg_and}
		Let $n \in \mathbb{N}$ and let $2^{-n} \le \epsilon \le \tfrac{1}{3}$. Then:
		
		\[ \epsadeg{\ANDn} = \Theta\left(\sqrt{n \cdot \log(\sfrac{1}{\epsilon})}\right) \]
	\end{theorem}
\end{leftbar}

Consider a Boolean function $f$. If the representing polynomial of $f$ has small $\ell_1$-norm, then one may use the following straightforward approach for constructing a low-degree approximating polynomial for $f$: approximate (with sufficiently small $\epsilon$) every \textit{monomial} of the representing polynomial. Since each monomial is an $\AND$ function, we may appeal to Theorem \ref{thm:epsilon_apx_deg_and} to obtain a low-degree approximation. Thus, the polynomial approximating $f$ is given by summing the approximating polynomials for each of its monomials. The details of this scheme are shown in the following lemma.

\begin{leftbar}
	\begin{lemma}
		\label{lemma:low_l1_norm_apx_deg}
		Let $f: \{0,1\}^n \rightarrow \{0,1\}$ be a Boolean function and let $p \in \mathbb{R}\left[x_1, \dots, x_n\right]$ be the unique multilinear polynomial representing $f$. If $2 < \norm{p}_1 < 2^{n}$, then:
		
		\[ \forall \epsilon \in \left(\frac{1}{\norm{p}_1}, \frac{1}{3}\right):\ \ \epsadeg{f} = \mathcal{O}\left(\sqrt{n \cdot \log \norm{p}_1} \right) \]
	\end{lemma}
\end{leftbar}
\begin{proof}
	Let $f: \{0,1\}^n \rightarrow \{0,1\}$ and let $p$ be the multilinear polynomial representing $f$, where:
	
	\[ p(x_1, \dots, x_n) = \sum_{S \subseteq [n]} a_S \cdot \ANDS(x_1, \dots, x_n) \]
	
	and $\ANDS(x_1, \dots, x_n) = \Pi_{i \in S} x_i$, i.e., the monomial corresponding to the set $S \subseteq [n]$. By Theorem \ref{thm:epsilon_apx_deg_and}, each monomial $\Pi_{i \in S} x_i$ can be approximated pointwise with error at most $\frac{\epsilon}{\norm{p}_1}$, by a multilinear polynomial $\widetilde{\ANDS}$ of degree $\mathcal{O}\left(\sqrt{n \cdot \log(\sfrac{\norm{p}_1}{\epsilon})}\right)$ = $\mathcal{O}\left(\sqrt{n \cdot \log \norm{p}_1} \right)$, using only the variables corresponding to $S$. Thus, consider the multilinear polynomial $\widetilde{p} \in \mathbb{R}\left[x_1, \dots, x_n\right]$:
	
	\[ \widetilde{p}(x_1, \dots, x_n) = \sum_{S \subseteq [n]} a_S \cdot \widetilde{\ANDS}(x_1, \dots, x_n) \]
	
	By construction, $\deg(\widetilde{p}) = \max \lbrace\deg\left(\widetilde{\ANDS}\right) : S \subseteq [n] \rbrace = \mathcal{O}(\sqrt{n \cdot \log {\norm{p}_1}})$. It remains to show that $\widetilde{p}$ $\epsilon$-approximates $f$. For any $x \in \{0,1\}^n$:
	
	\begin{equation*}
	\begin{split}
	\left|f(x_1, \dots, x_n) - \widetilde{p}(x_1, \dots, x_n)\right| &= \left|p(x_1, \dots, x_n) - \widetilde{p}(x_1, \dots, x_n)\right| \\
	&= \left|\sum_{S \subseteq [n]} a_S \cdot \left(\ANDS(x_1, \dots, x_n) -
	\widetilde{\ANDS}(x_1, \dots, x_n)\right)\right| \\
	&\le \sum_{S \subseteq [n]} |a_S| \cdot \left|\ANDS(x_1, \dots, x_n) - \widetilde{\ANDS}(x_1, \dots, x_n)\right| \\
	&\le \frac{\epsilon}{\norm{p}_1} \cdot \sum_{S\subseteq [n]} |a_S| \le \epsilon \qedhere
	\end{split}
	\end{equation*}
\end{proof}

The proof of Theorem \ref{thm:upper_bound_apx_deg} now follows.

\begin{proof}
	Let $n > 1$ and let $2^{-n \log n} \le \epsilon \le \tfrac{1}{3}$. Using Corollary \ref{cor:bpm_n_star_low_l1_norm}, Lemma \ref{lemma:apx_deg_dual} and Lemma \ref{lemma:low_l1_norm_apx_deg}, we obtain:
	
	\[ \epsadeg{\BPMn} = \epsadeg{\BPMnstar} = \mathcal{O}(n^{\sfrac{3}{2}} \sqrt{\log n}) \qedhere \]
\end{proof}
\section{The Lower Bound on $\adeg{\BPMn}$}

In this section we obtain a lower bound on the approximate degree of perfect matching, which matches the upper bound of Theorem \ref{thm:upper_bound_apx_deg} up to the low order term $\sqrt{\log n}$. Our lower bound applies to the constant-error approximate degree.

\begin{leftbar}
	\begin{theorem}
		\label{thm:lower_bound_apx_deg}
		Let $n > 2$. The approximate degree of $\BPMn$ is bounded by:
		
		\[ \adeg{\BPMn} = \Omega(n^{\sfrac{3}{2}}) \]
	\end{theorem}
\end{leftbar}

To this end, we begin with a brief ``warmup''; recounting some known lower bound techniques and their shortcomings when applied to the analysis of $\BPMn$. These techniques will only allow us to work our way up to a linear bound of $\Omega(n)$. To obtain the bound $\Omega(n^{\sfrac{3}{2}})$, we rely on a recent and powerful Theorem, due to Aaronson, Ben-David, Kothari, Rao and Tal \cite{aaronson2020degree}, which was inspired by Huang's proof of the Sensitivity Conjecture \cite{huang2019induced}. 

\subsection{A Brief Warmup: $\adeg{\BPMn} = \Omega(n)$}

\paragraph{Symmetrization Arguments} One of the best understood families of Boolean functions, in the context of approximate degree, is that of \textit{Symmetric Functions} -- functions which are invariant to any permutation of the input bits (ergo, depend only on the Hamming weight). Symmetric functions are simple in the following sense: their behaviour can be fully captured by a univariate polynomial, whose formal variable represents the Hamming weight. It is well known that averaging a multivariate polynomial over all orbits under the action of the Symmetric group to produce a univariate polynomial, can be performed in a manner that does not increase the degree. This technique, and related ones, are often referred to as ``Symmetrization'' (see, e.g., \cite{minsky1988perceptrons, aaronson2019quantum}). 

Symmetrization arguments have classically been applied to symmetric functions (\cite{nisan1994degree, paturi1992degree}), but have also found their use in the analysis of certain non-symmetric functions, including halfspaces \cite{sherstov2013intersection} and the $\AND$-$\OR$ tree \cite{kretschmer2021lower}. Given a univariate polynomial, one can appeal to Markov-Bernstein-type inequalities which roughly state that a bounded univariate polynomial with high (first, or higher order) derivative, must have sufficiently large degree. This is the general framework through which lower bounds on approximate degree are shown via symmetrization.

Could a similar approach be applied to $\BPMn$? Let us recall that the bipartite perfect matching function is \textit{monotone} -- if $G$ has a perfect matching, then so does any graph $H \supseteq G$ (i.e., the property is not diminished by the addition of edges). Like all monotone graph properties \cite{friedgut1996every}, bipartite perfect matching exhibits a sharp threshold. Erd\H{o}s and R\'enyi \cite{erdos1964random} first considered the probability of a perfect matching occuring in the random graph model $G(n, n,p)$, when $p = \frac{f(n) + \ln n}{n}$:

\begin{equation*}
	\begin{split}
		\lim_{n \rightarrow \infty}\ \  \Pr_{G \sim G\left(n,n,p\right)}\left[G \text{ has a P.M.}\right] = \begin{cases*}
							0 & $f(n) \rightarrow -\infty$ \\
							e^{-2e^{-c}} & $f(n) \rightarrow c$ \\
							1 & $f(n) \rightarrow \infty$
					   \end{cases*}
	\end{split}
\end{equation*} 

Thus, $\BPMn$ exhibits a sharp threshold at $n \ln n$. Paturi \cite{paturi1992degree} proved that the $k$-threshold function over $n$ input bits, which is the Boolean function $\mathbbm{1}\{|x| \ge t\}$, has approximate degree $\Theta(\sqrt{t \cdot (n-t+1)})$. Had the width of $\BPMn$'s threshold (the ``critical window'') been bounded by a constant, a similar argument would have implied that $\adeg{\BPMn} = \Omega(n^{\sfrac{3}{2}} \sqrt{\log n})$, exactly recovering our upper bound. Nevertheless, since the critical window for matching has width $\Theta(n)$ (as exhibited above), applying the same analysis would only yield a rather weak bound, on the order of $\tilde{\Omega}(\sqrt{n})$.

\paragraph{Monotonicity and Sensitivity} Let us briefly recall two well known results. The first result \cite{nisan1991crew} states that the \textit{block sensitivity} and \textit{sensitivity} are identical, for all \textit{monotone} Boolean functions. The second relates block sensitivity to approximate degree \cite{nisan1994degree}: $\sqrt{6} \cdot \adeg{f} \ge \sqrt{\bs(f)}$, for all Boolean functions $f$. Consequently, since $\BPMn$ is monotone, it suffices to bound its sensitivity.

Note that the sensitivity of $\BPMn$ at any \textit{elementary} graph is zero. Indeed, consider an elementary graph $G$. By construction, $G$ has a perfect matching. Using Hetyei's characterization of elementary graphs (Theorem \ref{thm:hetyei_conditions}), we know that any strict subset of $G$'s left vertices has strictly positive surplus -- thus removing any single edge of $G$ cannot violate Hall's condition. In fact, the same argument applies to any matching covered graph whose connected components are 2-connected (i.e., no component is $K_2$). Since asymptotically almost all bipartite balanced graphs are elementary (by a simple probabilistic method argument, see \cite{beniamini2020bipartite}), we immediately deduce that the \textit{average sensitivity} of $\BPMn$ is exponentially small.

Nevertheless, to obtain the lower bound it suffices to exhibit a single sensitive input. The following proposition constructs such an input.

\begin{leftbar}
	\begin{proposition}
		\label{prop:max_sens}
		Let $n > 1$. The sensitivity of $\BPMn$ is at least: 
		
		\[ \sens(\BPMn) \ge \begin{dcases*}
			\frac{n}{2} \left(\frac{n}{2} + 1\right),& $n \equiv 0 \pmod 2$ \\
			\left(\frac{n-1}{2} + 1\right)^2,& $n \equiv 1 \pmod 2$
		\end{dcases*} \]
	\end{proposition}
\end{leftbar}
\begin{proof}
	Without loss of generality, we provide a proof for the case $n \equiv 0 \pmod 2$. Let $k = \frac{n}{2}$ and let $A = \{a_1, \dots, a_{2k}\}$, $B = \{b_1, \dots, b_{2k}\}$ be two sets. Let $G$ be the graph composed of the following two disjoint paths: $P_1 = (a_1, b_1, a_2, b_2, \dots, a_k, b_k, a_{k+1})$, $P_2 = (b_{k+1}, a_{k+2} \dots, a_{2k}, b_{2k})$.
	
	$P_1$ is a connected component over an odd number of vertices, thus $G$ has no perfect matching. However, $\forall a_i \in P_1,\ \forall b_j \in P_2$, by adding the edge $(a_i, b_j)$ we obtain a graph that \textit{does} have a perfect matching, since taking the edge $(a_i,b_j)$ would split each path into two even length paths (or one path, if $a_i$ or $b_j$ are extremal vertices in $P_1$, $P_2$), each of which have a perfect matching. Consequently, the sensitivity at $G$ is at least $k(k+1) = \frac{n}{2}\left(\frac{n}{2} + 1\right) = \Theta(n^2)$.
\end{proof}

\begin{leftbar}
	\begin{corollary}
		$\forall n \in \mathbb{N}:\ \adeg{\BPMn} = \Omega(n)$.
	\end{corollary}
\end{leftbar}

\subsection{Obtaining the Bound $\adeg{\BPMn} = \Omega(n^{\sfrac{3}{2}})$}

Aaronson, Ben-David, Kothari, Rao and Tal \cite{aaronson2020degree} recently proved that for any \textit{total} Boolean function $f$, $\deg(f) = \mathcal{O}\left(\adeg{f}^2\right)$ (which is optimal, as exemplified by the $\ORn$ function). Their proof involves two main steps. First, they make the key observation that, at the heart of Huang's proof for the sensitivity conjecture \cite{huang2019induced}, there (implicitly) lies a new complexity measure: \textbf{Spectral Sensitivity}. Then, their main technical Theorem shows the aforementioned quantity lower-bounds approximate degree. It is this relation that we wish to leverage. \footnote{From the quadratic relation between degree and approximate degree \cite{aaronson2020degree} and using the fact that $\BPMn$ has full degree \cite{beniamini2020bipartite}, the \textit{weaker} lower bound of $\adeg{\BPMn} = \Omega(n)$ also immediately follows.}

\subsubsection{Spectral Sensitivity and the Sensitivity Graph}

\begin{leftbar}
	\begin{definition}(Sensitivity Graph \cite{aaronson2020degree})
		\label{def:sens_graph}
		Let $f: \{0,1\}^n \rightarrow \{0,1\}$ be a Boolean function. The \textbf{Sensitivity Graph} of $f$ is the graph $G_f$ over the vertices  $\{0,1\}^n$ whose edges are defined by:
		\begin{equation*}
			\begin{split}
				\forall x,y \in \{0,1\}^n:\ \{x,y\} \in E(G_f) \iff |x \oplus y| = 1 \land f(x) \ne f(y)
			\end{split}
		\end{equation*}

		Thus, $G_f$ is the subgraph containing all the bi-chromatic edges of the $n$-dimensional Hypercube whose vertices are labeled by $f$ (the ``$f$-cut'' of the Hypercube).	
	\end{definition}
\end{leftbar}

\begin{leftbar}
	\begin{definition}(Spectral Sensitivity \cite{aaronson2020degree})
		\label{def:spectral_sensitivity}
		Let $f: \{0,1\}^n \rightarrow \{0,1\}$ be a Boolean function and let $G_f$ be its sensitivity graph. The \textbf{Spectral Sensitivity} of $f$ is defined by $\lambda(f) \eqdef \rho(G_f)$.
	\end{definition}
\end{leftbar}

Observe that the sensitivity graph of \textit{any} Boolean function $f$ is a bipartite graph whose bipartitions are given by $f^{-1}(0)$ (hereafter, the ``left'' vertices) and $f^{-1}(1)$ (the ``right'' vertices). Clearly as all the edges of the sensitivity graph are bi-chromatic, these two sets form a valid bipartition. In the case of $\BPMn$, we note that (perhaps rather confusingly) the sensitivity graph is a bipartite graph in which each vertex $x \in \{0,1\}^{n^2}$ is, itself, associated with a bipartite graph (corresponding to the input $x$).

Under this notation, the main Theorem of \cite{aaronson2020degree} states the following.
\begin{leftbar}
	\begin{theorem}(\cite{aaronson2020degree})
		\label{thm:spec_sens_lowerbound}
		For any total Boolean function $f: \{0,1\}^n \rightarrow \{0,1\}$, we have:
		\[ \lambda(f) = \mathcal{O}(\adeg{f})\]
	\end{theorem}
\end{leftbar}

\subsubsection{A Tight Bound on the Spectral Sensitivity of $\BPMn$}

In what follows, we obtain matching upper and lower bounds on the Spectral Sensitivity of $\BPMn$.

\begin{leftbar}
	\begin{theorem}
		\label{thm:bound_spec_sens_bpm_n}
		For any $n \in \mathbb{N}$, we have:
		\[ \lambda(\BPMn) = \Theta(n^{\sfrac{3}{2}}) \]
	\end{theorem}
\end{leftbar}

This tight bound on $\lambda(\BPMn)$ yields our approximate degree lower bound, and also shows that this is the \textit{best bound attainable} by the method of Spectral Sensitivity for the perfect matching function.

\begin{leftbar}
	\begin{corollary}
		\label{cor:bpm_apx_deg_lower_bound}
		For any $n \in \mathbb{N}$, we have:
		\[ \adeg{\BPMn} = \Omega(n^{\sfrac{3}{2}}) \]
	\end{corollary}
\end{leftbar}
\begin{proof}
	Follows from Theorem \ref{thm:bound_spec_sens_bpm_n} and Theorem \ref{thm:spec_sens_lowerbound}.
\end{proof}

\subsubsection{The Upper Bound $\lambda(\BPMn) = \mathcal{O}(n^{\sfrac{3}{2}})$}

The spectrum of bipartite graphs has several nice properties. Their eigenfunctions come in pairs with negated eigenvalues (thus their spectrum is symmetric). Another well known result regarding the spectrum of bipartite graphs is H\"older's inequality for matrix norms:

\begin{leftbar}
	\begin{proposition}
		\label{prop:max_spec_radius_bipartite}
		Let $G$ be a bipartite graph and let $\Delta_L$ and $\Delta_R$ be the maximal left and right degrees, correspondingly. Then:
		\[\rho(G) \le \sqrt{\Delta_L \Delta_R}\]
	\end{proposition}
\end{leftbar}

Recall that the degree of any vertex in the sensitivity graph is equal, by definition, to the number of bi-chromatic edges incident to it -- which is its sensitivity. Thus, by Proposition \ref{prop:max_spec_radius_bipartite}, for any Boolean function $f$ we have $\lambda(f) \le \sqrt{s_0(f) \cdot s_1(f)}$, where $s_b(f) \eqdef \max_{x \in f^{-1}(b)} \sens_f(x)$, $\forall b \in \{0,1\}$. This simple observation suffices to obtain the upper bound:

\begin{leftbar}
	\begin{proposition}
		For any $n \in \mathbb{N}$, we have:
		\[\lambda(\BPMn) \le n^{\sfrac{3}{2}} \]
	\end{proposition}
\end{leftbar}
\begin{proof}
	Recall that $\lambda(\BPMn) \le \sqrt{s_0(\BPMn) \cdot s_1(\BPMn)}$. Since the sensitivity of \textit{any} input is at most $n^2$ (the number of input bits), we immediately have $s_0(\BPMn) \le n^2$ (and in fact, by Proposition \ref{prop:max_sens}, $s_0(\BPMn) = \Theta(n^2)$). As for the $1$-sensitivity, clearly for every $G \in \BPM_n^{-1}(1)$, the only sensitive edges are those present in some matching. The union of all matchings is matching-covered, and every edge in an elementary component is sensitive if and only if the component is $K_2$ (see Theorem \ref{thm:hetyei_conditions}), thus $s_1(\BPMn) \le n$.
	
%
\end{proof}

\subsubsection{The Lower Bound $\lambda(\BPMn) = \Omega(n^{\sfrac{3}{2}})$}

Let us now consider the lower bound. By Cauchy's Interlace Theorem (and using the fact that for bipartite graphs, $\rho(G) = \lambda_1$), it holds that for any \textit{bipartite} graph $G$, the spectral radius of $G$ is no smaller than that of any \textit{induced subgraph} of $G$. Thus, it suffices to exhibit an induced subgraph of the sensitivity graph of $\BPMn$, whose spectral radius is large. In the following Theorem, we construct such a connected bi-regular induced subgraph, and bound its spectral radius.

\begin{leftbar}
	\begin{theorem}
		\label{thm:min_spec_radius_bipartite}
		For any $n > 2$ we have:
		\[\lambda(\BPMn) \ge \frac{n^{\sfrac{3}{2}}}{3 \sqrt{3}} - \mathcal{O}(n) = \Omega(n^{\sfrac{3}{2}}) \]
	\end{theorem}
\end{leftbar}
\begin{proof}
	Let $n > 2$ and let $\frac{n + 1}{3} < k < n$ be a natural number. Let $A = A_1 \cupdot A_2$ and $B = B_1 \cupdot B_2$ be disjoint sets such that $|A_1| = |B_1| = k$ and $|A_2| = |B_2| = n-k$. For every $t \in \mathbb{N}^{+}$, denote the set of all matchings joining $t$ vertices of $A_2$ with $t$ vertices of $B_2$, by $M_t(A_2, B_2)$. Consider the following two sets of graphs:
	
	\begin{equation*}
		\begin{split}
			\mathcal{R} &= \left\{ G=(A \cupdot B,(A_1 \times B_2) \cupdot (A_2 \times B_1) \cupdot E(M))\ :\ M \in M_{n-2k}(A_2, B_2)\right\} \\
			\mathcal{L} &= \left\{ G=(A \cupdot B,(A_1 \times B_2) \cupdot (A_2 \times B_1) \cupdot E(M))\ :\ M \in M_{n-2k-1}(A_2, B_2)\right\}
		\end{split}
	\end{equation*}
	
	Let $G_{\BPMn}$ be the sensitivity graph of $\BPMn$, and let $H = G_{\BPMn}\left[\mathcal{L} \cupdot \mathcal{R}\right]$ be its induced subgraph over the aforementioned set of graphs (vertices). By Cauchy's Interlace Theorem the spectral radius of $H$ is at most that of $G$, therefore:
	
	\[ \lambda(\BPMn) = \rho(G_{\BPMn}) \ge \rho(H) \]
	
	\begin{figure}[h!]
		\centering
		\includegraphics[height=8.5cm]{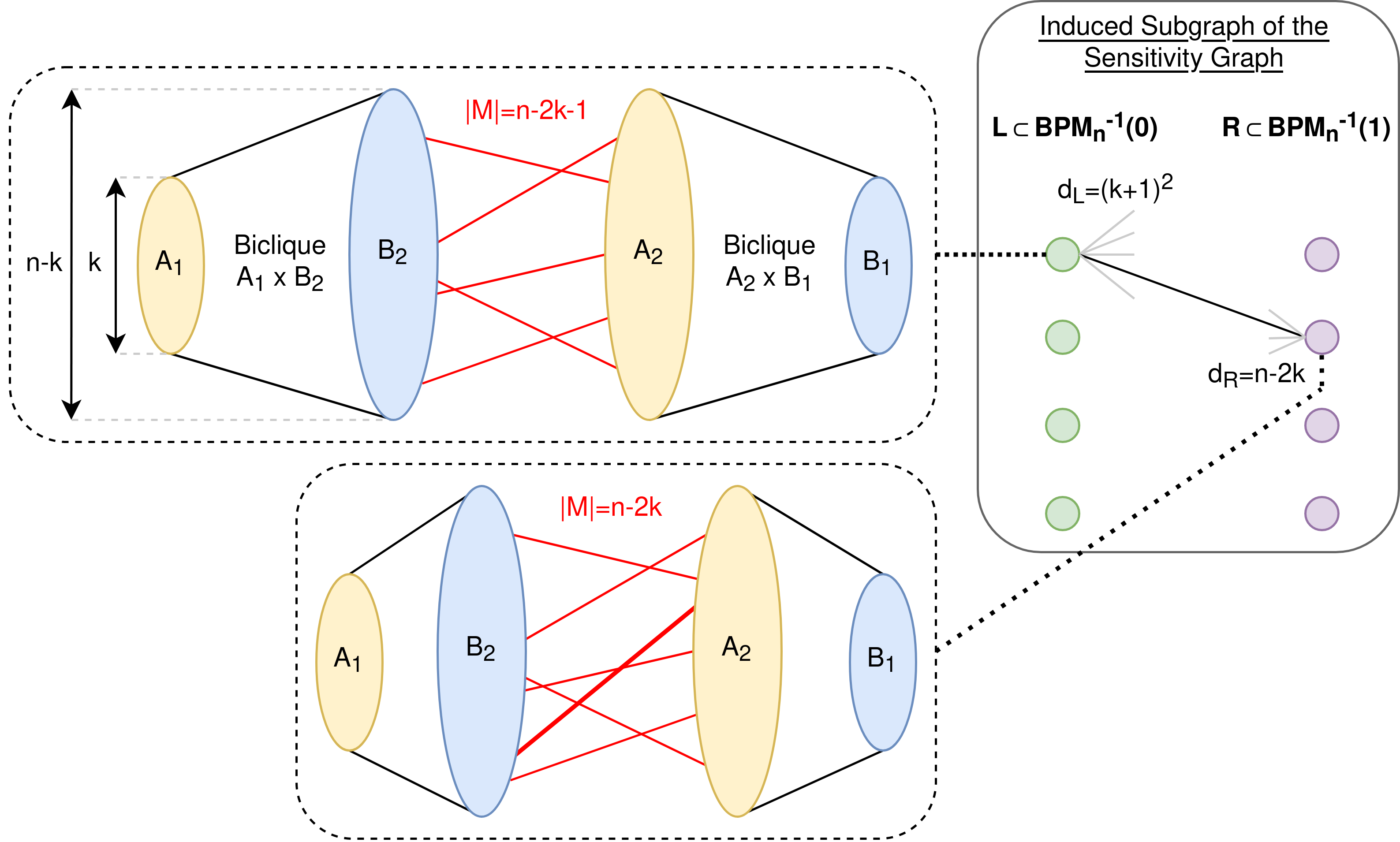}
		\caption{The induced subgraph $H = G_{\BPMn}\left[\mathcal{L} \cupdot \mathcal{R}\right]$ of the sensitivity graph for $\BPMn$.}
	\end{figure}
	
	Observe that every graph $G \in \mathcal{R}$ has a perfect matching, which can be constructed by taking the $(n-2k)$-size matching between $A_2$ and $B_2$ and matching the remaining $k$ vertices of $A_2$ with $B_1$, and similarly the remaining $k$ vertices of $B_2$ with $A_1$ (this can always be done, since the bicliques $K_{A_2,B_1}$, $K_{A_1,B_2}$ are subgraphs of $G$). Conversely, every graph $G \in \mathcal{L}$ does not have a perfect matching. For example, the set $A_2$ violates Hall's condition, since: $|N(A_2)| = n-2k-1 + k = n - k - 1 < n - k = |A_2|$. Thus, $\mathcal{R}$ is the right bipartition of $H$, and $\mathcal{L}$ is its left bipartition.
	
	Let us characterize the edges of $H$. Let $G \in \mathcal{R}$ and let $M$ be its corresponding $(n-2k)$-size matching between $A_2$ and $B_2$. For every edge $e \in M$, we have by construction $(G \setminus \{e\}) \in \mathcal{L}$. Furthermore, for any edge $e \in (E(G) \setminus M)$, the graph $(G \setminus \{e\})$ does not contain one of the bicliques $K_{A_2,B_1}$, $K_{A_1,B_2}$, and is therefore not in $\mathcal{R}$. Consequently the degree of each right vertex of $H$ is:
	
	\[ d_R \eqdef \deg_H(G) = |M| = n-2k\]
	
	Similarly, let $G \in \mathcal{L}$ and let $M$ be its $(n-2k-1)$-size matching between $A_2$ and $B_2$. Denote by $S$, $T$ the left and right vertices of $M$, correspondingly. Then, for any $u \in (A_2 \setminus S)$, $v \in (B_2 \setminus T)$, the graph $G \cupdot \{(u,v)\}$ has a $(n-2k)$-size matching, and is thus in $\mathcal{R}$. Adding any other edge $e$ to $G$ would either join a vertex from $A_1$ to a vertex from $B_1$, or $e$ would be incident to a vertex in $M$. In both cases, $G \cupdot \{e\}$ is not in $\mathcal{L}$. Thus the degree of each left vertex of $H$ is:
	
	\[ d_L \eqdef \deg_H(G) = (|A_2| - |S|) \cdot (|B_2|-|T|) = (k+1)^2\]

	Finally, observe that any bi-regular bipartite graph, and in particular $H$, satisfies $\rho(H) \ge \sqrt{d_L \cdot d_R}$ (this follows, for example, by considering the eigenfunction which places weight $\sqrt{d_L}$ on each left vertex, and $\sqrt{d_R}$ on each right vertex)\footnote{We remark that it is not hard to see that $H$ is connected for any $n>2$. Thus the top eigenvalue of $H$ is \textit{simple}, and consequently our bound does not freely extend, by interlacing, to eigenvalues other than the spectral radius of $G_{\BPMn}$.}. To conclude the proof, fix $k = \lfloor\frac{n}{3}\rfloor - 1$. Thus:
	
	\begin{equation*}
		\begin{split}
			\lambda(\BPMn) &= \rho(G) \ge \rho(H) \ge \sqrt{ \left(\left\lfloor\frac{n}{3}\right\rfloor\right)^2 \cdot \left(n - 2\left(\left\lfloor\frac{n}{3}\right\rfloor - 1\right)\right)} = \frac{n^{\sfrac{3}{2}}}{3 \sqrt{3}} - \mathcal{O}(n) \qedhere
		\end{split}
	\end{equation*}		
\end{proof}
\section{Towards Fine Grained Bounds for Bipartite Perfect Matching}
\label{section:fine_grained_bounds}

The main thrust of this section, and indeed one of the motivating factors for the work in this paper, revolves around the following longstanding open question:

\savenotes
\begin{leftbar}
\begin{openprob} (The $n^{\sfrac{5}{2}}$-Barrier for Bipartite Matching\footnote{In fact we are only interested in \textit{polynomial} improvements to this running time, i.e., bounds of the form $n^{5/2 - \varepsilon}$ for some constant $\varepsilon > 0$.})
	\label{openprob:n52barrier}
	\begin{center}
		Does there exist a \textbf{deterministic} algorithm for bipartite perfect matching running in time $o(n^{\sfrac{5}{2}})$?
	\end{center}
\end{openprob}
\end{leftbar}
\spewnotes

Hopcroft and Karp's \cite{hopcroft1973n} algorithm, designed half a century ago, attains a runtime of $\mathcal{O}\left(n^{\sfrac{5}{2}}\right)$ when applied to \textit{dense} graphs (i.e., when the number of edges is $\Theta(n^2)$). Since then, no known deterministic algorithm has been able to break this barrier, in the dense regime. To make matters concrete, in what follows let us consider the \textit{decision variant} of the problem, as represented by $\BPMn$; we are given a balanced bipartite graph with $n$ vertices in each bipartition, and wish to determine whether a perfect matching exists. Secondly, let us fix the following computational model.

\paragraph{The Demand Query Model.} In recent work, Nisan \cite{nisan2021demand} introduced a new \textit{concrete complexity model} for bipartite matching, known as the ``Demand Query Model''. This model appears to be particularly well-suited for the matching problem, for two primary reasons. Firstly, Nisan showed that \textit{combinatorial} matching algorithms can be efficiently simulated within the model (in fact, this holds even for parallel, online, approximate and other classes of algorithms, see \cite{nisan2021demand}). For instance, Hopcroft and Karp's algorithm, whose running time is $\mathcal{O}(n^{\sfrac{5}{2}})$, can be ``translated'' into demand query algorithm making $\mathcal{O}(n^{\sfrac{3}{2}})$ queries. Since each query can be trivially simulated in $\mathcal{O}(n)$ time, this appears to capture the complexity of the aforementioned algorithm in a fine-grained manner. Secondly, the queries in this model are \textit{simple enough} that \textit{we could hope to prove lower bounds against them}.

In this framework, algorithms are modeled by decision trees. Each internal node corresponds to a demand query, and each leaf is labeled by an output, either $0$ or $1$. A demand query consists of a left vertex $u$ and an ordering $\pi \in S_n$, induced on the right vertices. The result of such a query is the first right vertex $v$, according to the ordering $\pi$, for which the edge $(u,v)$ exists in the graph (or $\bot$ if no such edge exists). A root-to-leaf path in the tree corresponds to a particular set of answers to the queries made along the path. Thus, the set of all such paths partitions the set of all graphs $G \subseteq K_{n,n}$, whereby each graph $G$ is associated with a single leaf. Any graph $G \subseteq K_{n,n}$ which is ``consistent'' with the answers made along a root-to-leaf path, must also be consistent with the labeling of that leaf. The ``cost'' of an algorithm in this model is measured by the depth of the tree (i.e., the worst-case amount of queries made on any particular input). As this is an information-theoretic model, we disregard the amount of computation necessary to construct (or deduce the existence of) a perfect matching, and instead only measure the minimal amount of \textit{information} required to do so.

\subsection{The Demand Query Complexity of Matching}

Open Problem \ref{openprob:n52barrier} remains as of yet unsettled. In light of the efficient simulation of combinatorial algorithms by the demand model, one could formulate the following closely related question: ``can one construct quasi-linear demand-query algorithms for matching?'', or in the contrapositive:

\begin{leftbar}
	\begin{openprob} (The Demand Query Complexity of Matching)
		\label{openprob:quasilinear_demand}
		\begin{center}
			Does there exist some constant $\varepsilon > 0$ such that $\demand(\BPMn) = \Omega(n^{1 + \varepsilon})$?
		\end{center}
	\end{openprob}
\end{leftbar}

\begin{figure}[h!]
	\centering
	\includegraphics[height=9.5cm]{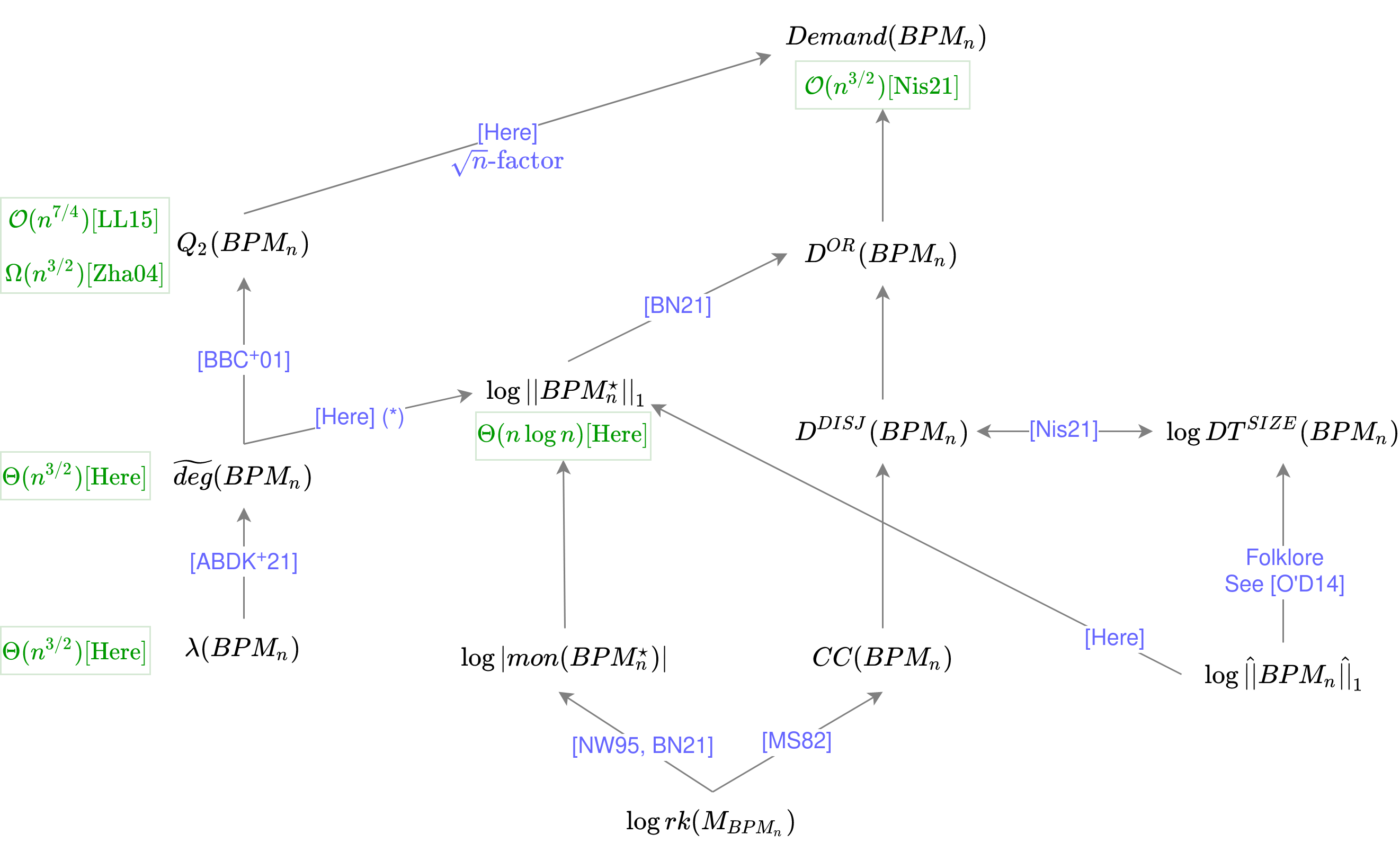}
	\caption{Relations between complexity measures of $\BPMn$. An arrow $f(n) \rightarrow g(n)$ indicates that $f = \widetilde{\mathcal{O}}(g)$ (i.e., excluding log factors) -- with two exceptions. The arrow $Q_2(\BPMn)$ $\rightarrow$ $\demand(\BPMn)$ incurs a $\sqrt{n}$-factor loss, see Subsection \ref{subsection:quantum_bounds}, and the arrow $\adeg{\BPMn} \rightarrow \log \norm{\BPMnstar}_1$ represents the bound provided by Lemma \ref{lemma:low_l1_norm_apx_deg}. Green blocks correspond to bounds on their adjacent quantity. Every arrow $f \rightarrow g$ is accompanied by the corresponding citation in blue, apart from trivial relations wherein citations are omitted. Bounds and relations marked by [Here] correspond to results shown in this paper. }
	\label{fig:complexity_measures}
\end{figure}

To better understand $\demand(\BPMn)$, we have drawn connections between the demand query complexity of $\BPMn$ and other, mostly algebraic, complexity measures relating to $\BPMn$ and its dual -- a representative collection of which are detailed in Figure \ref{fig:complexity_measures}.

The following table details the complexity measures appearing in Figure \ref{fig:complexity_measures}.

\renewcommand*\arraystretch{1.2}
\begin{table} [!htb]
\centering
\begin{tabular}{ |p{3cm}||p{11cm}|  }
	\hline
	\multicolumn{2}{|c|}{\textbf{Query Complexity Measures}} \\
	\hline
	\multicolumn{1}{|c||}{\textbf{Measure}} & \multicolumn{1}{|c|}{\textbf{Definition}} \\
	\hline
	$\demand(\BPMn)$ & The least depth of a decision tree computing $\BPMn$, whose internal nodes are labeled by demand queries. \\
	\hline
	$D^{\OR}(\BPMn)$ &  The least depth of a decision tree computing $\BPMn$, whose internal nodes are labeled by $\OR$s over arbitrary subsets of the input bits. \\
	\hline
	$D^{\DISJ}(\BPMn)$ & The least depth of a decision tree computing $\BPMn$, whose internal nodes are labeled by \textit{disjunctions} over \textit{literals}, e.g. $(x_1 \lor \bar{x}_3 \lor x_7)$. \\
	\hline
	$DT^{SIZE}(\BPMn)$ & The least amount of leaves in a \textit{classical} decision tree computing $\BPMn$. \\
	\hline
	$Q_2(\BPMn)$ & The bounded-error quantum query complexity of $\BPMn$. \\
	\hline
\end{tabular}
\end{table}
\begin{table} [!htb]
	\centering
	\begin{tabular}{ |p{3cm}||p{11cm}|  }
		\hline
		\multicolumn{2}{|c|}{\textbf{Communication Complexity Measures}} \\
		\hline
		\multicolumn{1}{|c||}{\textbf{Measure}} & \multicolumn{1}{|c|}{\textbf{Definition}} \\
		\hline
		$CC(\BPMn)$ & The two-party deterministic communication complexity of $\BPMn$, where we fix an \textit{arbitrary} partition over the input bits. \\
		\hline
		$rk(M_{\BPMn})$ & The real rank of the communication matrix corresponding to the above communication problem. \\
		\hline
	\end{tabular}
\end{table}
\begin{table} [!htb]
	\centering
	\begin{tabular}{ |p{3cm}||p{11cm}|  }
		\hline
		\multicolumn{2}{|c|}{\textbf{Algebraic Complexity Measures}} \\
		\hline
		\multicolumn{1}{|c||}{\textbf{Measure}} & \multicolumn{1}{|c|}{\textbf{Definition}} \\
		\hline
		$|mon(\BPMnstar)|$ & Number of non-zero coefficients in the unique representing polynomial. \\
		\hline
		$\norm{\BPMnstar}_1$ & Sum of magnitudes of coefficients in the unique representing polynomial. \\
		\hline 
		$\lambda(\BPMn)$ & The spectral sensitivity of $\BPMn$ (see Definition \ref{def:spectral_sensitivity}). \\
		\hline 
		$\adeg{\BPMn}$ & The approximate degree of $\BPMn$. \\
		\hline
	\end{tabular}
\end{table}

\subsection{Drawing the Connections} 

\paragraph{Decision Tree Measures.} It is not hard to see that every demand query can be simulated by at most logarithmically many $\OR$-queries, by performing binary search on the right vertices.  Similarly trivially, every $\OR$ query can be seen as a disjunction wherein no literal is negated, thus we also have $D^{\DISJ}(\BPMn) \le D^{\OR}(\BPMn)$. The latter quantity, $D^{\DISJ}(\BPMn)$, is of particular interest -- Nisan observed \cite{nisan2021demand} that for any Boolean function the least \textit{depth} of a \textit{disjunction decision tree} computing the function is equivalent, up to a $\log n$-factor, to the minimum \textit{size} of a \textit{classical decision tree} computing it. The minimum decision tree size computing a Boolean function is known to be related to Fourier-analytic properties of the function. For example, a folklore result states that it is lower bounded by the Fourier $\ell_1$-norm of the function (see e.g. \cite{o2014analysis}).

\paragraph{Communication Complexity Measures.} Given a disjunction decision tree computing a Boolean function, one naturally obtains a corresponding 2-party deterministic communication protocol. The protocol simply simulates the tree by ``solving'', at every step, the current disjunction. This simulation can be done efficiently, since any disjunction requires only $2$-bits of communication (Alice and Bob compute their parts of the disjunction separately, and communicate the answer bits to one another). In the argument above, the actual partition determining Alice and Bob's shares of the input bits is inconsequential. For any such fixed partition, one can consider the \textit{communication matrix}, which is the Boolean matrix whose rows are indexed by Alice's inputs, and columns by Bob's inputs. It is well known (by a result of \cite{mehlhorn1982vegas}), that the log of the real rank of this matrix yields a lower bound on the deterministic communication complexity of its corresponding problem.

\paragraph{The $\ell_1$-norm of $\BPMnstar$.} A surprisingly pivotal complexity measure arising in Figure \ref{fig:complexity_measures} is the $\ell_1$-norm of the dual function of matching. Firstly, this measure trivially bounds the number of \textit{monomials} appearing in its representing polynomial (since all coefficients are integers), which in turn bounds the rank of the communication matrix, by a classical result of \cite{nisan1995rank} (every monomial corresponds to a rank-1 matrix). The same quantity, $\norm{\BPMnstar}_1$, also bounds the Fourier $\ell_1$-norm of $\BPMn$ (equivalently $\BPMnstar$), as we observe in Corollary \ref{cor:bpm_n_star_low_l1_norm}, as well as yielding bounds on the approximate degree, via the scheme detailed in Lemma \ref{lemma:low_l1_norm_apx_deg}. With regards to lower bounds, in \cite{beniamini2020bipartite} it was shown that for any Boolean function $f$ it holds that $\log \norm{f^\star}_1$ lower bounds the least depth of an $OR$ decision tree computing $f$. 

Through our complete characterization of the dual polynomial given in Theorem \ref{thm:bpm_star_poly}, we were able to deduce the tight bound $\log \norm{\BPMnstar}_1 = \Theta(n \log n)$ (see Corollary \ref{cor:bpm_n_star_low_l1_norm}), thereby implying all of the aforementioned bounds. We conjecture that this low-norm representation of $\BPMnstar$ has more far-reaching consequences -- in particular, that it can be used to construct a quasi-linear deterministic communication protocol for the bipartite matching problem.\footnote{Indeed, it is not hard to show that for any \textit{monotone} Boolean function $f: \{0,1\}^n \rightarrow \{0,1\}$ and any partition over its inputs, we have $CC(f) \le \min\left\{|mon(f)|, |mon(f^\star)|\right\}^2$, which can be seen as a single step towards this direction.}

\paragraph{Approximate Degree and Quantum Query Complexity.} The ``polynomial method'' in quantum computation \cite{beals2001quantum} states that the acceptance probability of any $d$ query quantum algorithm can be written as a degree $2d$ polynomial. Thus, the approximate degree of any Boolean function serves as a lower bound on its Quantum query complexity. In this paper we have obtained tight upper and lower bounds on this quantity, showing that $\adeg{\BPMn} = \widetilde{\Theta}(n^{\sfrac{3}{2}})$. To complete the connections specified in Figure \ref{fig:complexity_measures}, it remains to relate the quantum query complexity to our main object of study; $\demand(\BPMn)$. 

\subsection{Quantum Bounds Imply Combinatorial Bounds}
\label{subsection:quantum_bounds}

In this section, we make one final simple observation regarding the demand query model: \textit{demand query algorithms can be efficiently simulated by quantum queries}. Recall that every demand query can be simulated by logarithmically many $\OR$-queries, each over at most $n$ bits (corresponding to the right vertices). In his seminal paper, Grover \cite{grover1996fast} showed that the $\ORn$ function can be computed, to constant error, using $\mathcal{O}(\sqrt{n})$ quantum queries (which is tight, see \cite{bennett1997strengths}). Thus, replacing each demand query by the majority over several invocations of Grover's algorithm, and using Chernoff's bound to suitably reduce the error, we obtain:\footnote{In fact, by a similar approach we can also show that for any Boolean function $f: \{0,1\}^n \rightarrow \{0,1\}$, the quantum query complexity is bounded by $Q_2(f) = \mathcal{O}\left(\sqrt{n} \cdot \log DT^{SIZE}(f) \cdot \log \log DT^{SIZE}(f)\right)$, where $DT^{SIZE}(f)$ is the minimal \textit{size} of a classical decision tree computing $f$. This observation might be useful in cases where there exist relatively ``unbalanced'' decision trees computing $f$.}
 
\begin{leftbar}
	\begin{proposition}
		If there exists a demand query algorithm for $\BPMn$ making at most $d$ queries, then:
			\[ Q_2(\BPMn) = \mathcal{O}(\sqrt{n} \cdot d \cdot \polylog(d)) \]
	\end{proposition}
\end{leftbar}

Consequently, any lower bound of the form $Q_2(\BPMn) = \Omega(n^{\sfrac{3}{2} + \varepsilon})$, for some constant $\varepsilon > 0$, would imply a (polynomially) super-linear lower bound on the demand query complexity of $\BPMn$, thereby resolving Open Question \ref{openprob:quasilinear_demand}. Such a result might suggest that quasi-linear \textit{combinatorial algorithms} for bipartite perfect matching are improbable, which we consider a very interesting prospect. Nevertheless, at present the quantum query complexity of $\BPMn$ remains undetermined. Lin and Lin \cite{lin2015upper} constructed an efficient quantum algorithm, yielding an upper bound of $\mathcal{O}\left(n^{\sfrac{7}{4}}\right)$. Conversely, through Ambainis' adversary technique, Zhang \cite{zhang2004power} has obtained an upper bound of $\Omega\left(n^{\sfrac{3}{2}}\right)$. Our main theorem (Theorem \ref{thm:apx_deg_bpm}) implies that this lower bound \textit{cannot be (polynomially) strengthened by the ``Polynomial Method''}. In fact, neither can Ambainis' adversary bounds be used to this end, since it is known (see e.g. \cite{zhang2004power}) that the best bound attainable by this method cannot exceed $\sqrt{C_0(f) C_1(f)}$. Closing this gap is left as an open problem.

\begin{leftbar}
	\begin{openprob} (Quantum Query Complexity of Matching) 
		\label{openprob:quantum_query_bpmn}
		Close the gap on $Q_2(\BPMn)$.
	\end{openprob}
\end{leftbar}

\section{Acknowledgments}

I would like to thank Noam Nisan and Nati Linial for helpful discussions. I would also like to thank Bruno Loff for pointing out typographical errors in an earlier version of this paper.

\bibliography{dual_polynomial}
\bibliographystyle{alpha}

\end{document}